\pdfoutput=1
\documentclass[aps,prA,amssymb,notitlepage ,superscriptaddress,nofootinbib,nopreprintnumbers]{revtex4-1}  

\usepackage{graphicx}
\usepackage{picture}
\usepackage{placeins}
\usepackage{float}
\DeclareGraphicsExtensions{.pdf,.eps} 
\usepackage{amsmath,scalefnt}
\usepackage{mathtools}
\usepackage{amssymb}
\usepackage{verbatim}
\usepackage{amsmath,amsfonts,bbm}
\usepackage{amsbsy}
\usepackage{color}
\usepackage{cancel}
\usepackage{soul}
\usepackage{dsfont}
\usepackage[svgnames]{xcolor}

\DeclareMathOperator{\csch}{csch}

\usepackage[breaklinks=true,colorlinks=true,linkcolor=DarkBlue,urlcolor=DarkBlue,citecolor=DarkBlue]{hyperref}

\usepackage{subfigure}

\usepackage{braket}

\newcommand{\da}[0]{^{\dagger}}

\def\bs#1{\boldsymbol{#1}}
\def\ba#1{\left(\begin{array}{#1}}
	\def\ea{\end{array}\right)}
\def\bsm{\left(\begin{smallmatrix}}
	\def\esm{\end{smallmatrix}\right)}

\def\bs#1{\boldsymbol{#1}}

\begin{document}

\title{Active Interferometry with Gaussian Channels}

\author{Richard Howl}
\email[Author to whom correspondence should be addressed: ]{richard.howl1@nottingham.ac.at}
  \affiliation{School of Mathematical Sciences,
	University of Nottingham,
	University Park,
	Nottingham NG7 2RD,
	United Kingdom}

\author{Ivette Fuentes}
  \affiliation{School of Mathematical Sciences,
University of Nottingham,
University Park,
Nottingham NG7 2RD,
United Kingdom}

\begin{abstract} 
We consider an interferometer that contains active elements, such as a parametric amplifier, with general two-mode Gaussian unitary channels rather than the usually considered phase-shift channel. We concentrate on a scheme based on the recently proposed pumped-up SU(1,1) active interferometer where all input particles participate in the parameter estimation, and from which a conventional SU(1,1)  interferometer is a limiting case. Using the covariance matrix formalism, we derive the quantum Fisher information of this active interferometer with a general two-mode Gaussian unitary channel, as well as the sensitivity for a number-sum measurement scheme, finding simple expressions for the latter. As an example application, we apply our results to Bose-Einstein condensates (BECs), and in particular a BEC gravitational-wave detector based on resonance, finding that the sensitivity of the detector can be improved  by several orders of magnitude with this new interferometry scheme.   
\end{abstract}
 
\maketitle

\section*{Introduction} \label{sec:Intro}

Since its invention by Michelson in the late 19$^{\rm{th}}$ century, interferometers have become a powerful tool for precision measurements, often achieving sensitivities that are not possible with any other known techniques. The most astonishing application is perhaps gravitational-wave (GW) detection, with the Laser Interferometer Gravitational-Wave Observatory (LIGO) using an interferometer of a very similar implementation to that designed by Michelson to observe GWs for the first time \cite{GWDetection}. The interferometers of LIGO are currently being upgraded to use techniques of quantum metrology, with the application of squeezed light expecting to improve the sensitivity, which has already demonstrated promising results in previous trials \cite{aasi2013enhanced}.   In general,  quantum correlations allow interferometers to go beyond the shot-noise limit $1/\sqrt{N}$, where $N$ is the number of probes, with Heisenberg scaling, $1/N$, considered the ultimate goal. One way to achieve this optimum scaling is to use an SU(2) interferometer design such as that implemented by LIGO but with squeezed input states \cite{Olivares2007}. However, it is also possible to  design an interferometer where the quantum correlations are generated within it, requiring fewer optical elements and making it more robust to losses. For example, an SU(1,1) (Mach-Zehnder) interferometer \cite{SU11}  is similar to an SU(2) Mach-Zehnder interferometer but with the   passive beam splitters replaced by active elements that parametrically create or annihilate  correlated particles. The interferometer, therefore, generates entanglement between its side modes, allowing for Heisenberg-scaling sensitivities. Such an interferometer has  been realized experimentally in various systems, including optical systems \cite{SU11Optical1,SU11Optical2}, hybrid atom-light systems  \cite{SU11AtomLight}, and spinor Bose-Einstein condensates (BECs) \cite{SU11BEC1,SU11BEC3}. 

However, generating a large number, $N$, of particles in the side modes is extremely challenging and the sensitivity is, therefore, easily beaten by  interferometers operating at the standard quantum limit with large input states, despite the poorer scaling with $N$. In order to overcome this issue, a variant of the SU(1,1) interferometer has recently been proposed  where the pump beam is mixed with the side modes such that all particles take part in the measurement \cite{PumpedUpSU11}. This essentially allows for $1 / \sqrt{N N_0}$ scaling where $N_0$ is the number of particles in the pump beam. Since, in general, $N_0 \gg N$, then this can improve the sensitivity of the original SU(1,1) interferometer  and, in particular, the sensitivity can never be worse \cite{PumpedUpSU11}.  This is similar to the fact that an SU(2) interferometer  with two squeezed input ports can provide the theoretically optimum sensitivity \cite{Olivares2007} but, in practice, it is currently preferable to use a large coherent beam and a squeezed state (see e.g.\ \cite{QuantumLimitsOptics}).

An  interferometer can be broken up into three stages: 1) the generation of the (reduced) quantum state of the side modes, 2) the quantum channel(s) that imprints the parameter to be estimated, and 3) the measurement process. The ultimate theoretical precision of the interferometer is obtained from the  saturation of the quantum Cram\'{e}r-Rao bound (QCRB) \cite{CramerRao1,CramerRao2}, $1 / \sqrt{M H}$, where $M$ is the number of repeated measurements and $H$ is the quantum Fisher information (QFI) introduced by Braunstein and Caves \cite{QFI} and optimizes over all possible measurement schemes such that it is  independent of stage 3). However, for a particular, measurement scheme, the  precision is bounded by the `classical' Cram\'{e}r-Rao bound, $1 / \sqrt{M F}$, where $F$ is referred to as the `classical' Fisher information for the measurement scheme, with $F \leq H$. This can also be related to the sensitivity of the interferometer $\Delta \epsilon$, defined by the minimum value of the shift in the parameter to be estimated $\epsilon$  that can
be sensed by considering changes in the values of the measured observable.

The second stage of an  interferometer is usually considered to consist of a phase-shift channel that imprints phases onto the two arms. Here, we extend this to general (two-mode) Gaussian unitary channels, i.e.\ mode-mixing and squeezing channels, and use the covariance matrix formalism (see Section \ref{sec:GaussianChannels}) to calculate the QFI and  sensitivity for a number-sum measurement scheme. In particular, we find straightforward expressions using the covariance matrix formalism, which was not used in \cite{PumpedUpSU11}, for calculating the sensitivity of general interferometers with number-sum measurements. This type of interferometer would be implementable in many systems, such as optical, hybrid atom-light, cold atoms and BECs. In particular, we apply our considered interferometry scheme to a GW detector that uses phonons of a BEC \cite{GWDetectorFirst}, finding that the sensitivity of the device can, in practice, be greatly improved in comparison to the previously considered scheme. In this case, all the active and passive elements of the interferometer would be applied to a single BEC in a trap and would not be separated in space as in a traditional interferometer, such as  Mach-Zehnder.

\section{Active interferometry}

The original active interferometer, the SU(1,1) interferometer, was proposed by Yurke \emph{et al.} \cite{SU11} and resembles a Mach-Zehnder interferometer but with the passive beam-splitters replaced by active elements.  The input state is a large coherent beam (the pump) that is sent to a parametric amplifier from which two beams, the side modes, are created in a two-mode squeezed vacuum state. The beams then undergo a phase shift after which they are recombined in a second parametric amplifier. Subsequently, a measurement is performed, such as a number-sum measurement, from which the phase shift can be estimated. The optimum sensitivity, $\Delta \phi$, of this interferometry is found to scale as $1/N$  \cite{SU11}. 

A recently proposed variant of the standard SU(1,1) interferometer, called pumped-up SU(1,1) \cite{PumpedUpSU11}, places a tritter (three-way beam splitter) between the first parametric amplifier and the phase-shift channel, and linearly mixes the side modes with the pump mode, which is assumed to be relatively undepleted after the fist parametric amplifier. This is illustrated in Figure \ref{fig:SU11}, with $\hat{U}_{\epsilon}$ representing the unitary transformation for a phase-shift channel, i.e.\  $\hat{U}_{\epsilon}= \exp (-i \phi \hat{N}/2)$ where $\phi := \phi_1 + \phi_2$, $\hat{N} := \hat{a}\da_1 \hat{a}_1 + \hat{a}\da_2 \hat{a}_2$, and $\hat{a}_1$ and $\hat{a}_2$ are the annihilation operators for the two side modes.\footnote{By the original definition \cite{SU11}, the pumped-up SU(1,1) interferometer \cite{PumpedUpSU11} is technically not an SU(1,1) interferometer since the unitary transformation of the tritter  does not belong to the SU(1,1) group. However, the original SU(1,1) (Mach-Zehnder) interferometer \cite{SU11} is derived when the tritter angle is zero.} Depending on the chosen angle, the tritter  can significantly increase the side mode particle numbers, at the expense of them no longer being in a squeezed vacuum state.  Since, the pump mode will have far more particles than the side modes after the first active element,  at least in foreseeable SU(1,1) interferometers,  this leads to an increase in the sensitivity of the interferometer.   Essentially, the sensitivity can now be made proportional to  $1 / \sqrt{N N_0}$ with $N_0 \gg N$ . 

\section{Active interferometry with Gaussian unitary channels} \label{sec:GaussianChannels}

\begin{figure*}[t!]
	\begin{center}
		\includegraphics[width=0.465\textwidth]{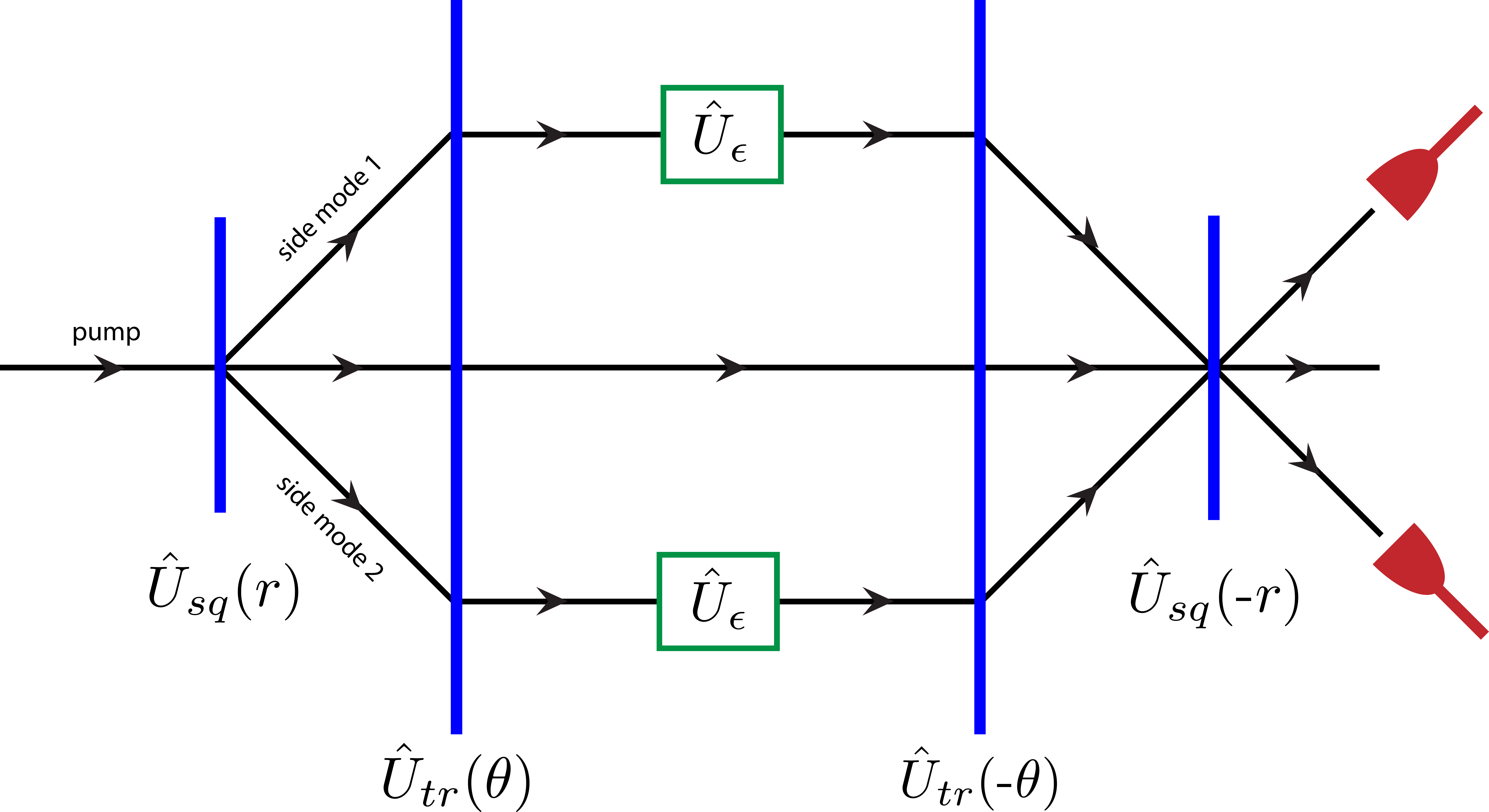}
	\end{center}
	\vspace{-0.5cm}
	\caption{Interferometer setup considered in Section \ref{sec:GaussianChannels}. The first stage is an active beam splitter $\hat{U}_{sq}(r)$, after which the side modes and pump are sent through a three-way passive beam splitter (a tritter) $\hat{U}_{tr} (\theta)$ as in the pumped-up SU(1,1) interferometer \cite{PumpedUpSU11}.  In pumped-up SU(1,1), the side modes then undergo a phase-shift channel $\hat{U}_{\epsilon} = U(\phi)$, whereas, for the scheme in Section \ref{sec:GaussianChannels}, the side modes  undergo a (two-mode)  Gaussian unitary channel, with  $\hat{U}_{\epsilon}$ representing the unitary transformation of the channel.  The beams are then recombined through the reverse tritter and active beam splitter operations.  Finally, a number-sum measurement is performed on the side modes.}
	\label{fig:SU11}
\end{figure*}

The schemes presented in the previous section assume that the side modes undergo unitary transformations that encode phases $\phi_1$ and $\phi_2$ on the respective states, and are sensitive to the total unitary transformation  $\hat{U}(\phi) = \exp (-i \phi \hat{N}/2)$  \cite{SU11}.  In this section,  we consider a pumped-up SU(1,1) interferometer but with the phase-shift channels now replaced with two-mode squeezing and mode-mixing channels, which is illustrated in Figure \ref{fig:SU11}, with $\hat{U}_{\epsilon}$ representing the unitary transformation for these channels. Together with the phase-shift channel, considered in \cite{PumpedUpSU11}, these form the complete set of unitary two-mode Gaussian channels (see e.g. \cite{CMF}).\footnote{For an analysis of the optimum input states for such channels using the QFI, see \cite{DomOptimum}.} They can also be considered as general (two-mode) Bogoliubov transformations, and  are generated through quadratic interaction Hamiltonians. Higher-order Hamiltonians will not, in general, be Gaussian-preserving \cite{CMF}. In this case, the side modes   undergo the following unitary transformations:
\begin{align} \label{eq:Us}
U(\xi)&= e^{\xi \hat{a}_1\da \hat{a}_2\da - \xi^{\ast} \hat{a}_1 \hat{a}_2}~~\mathrm{or}\\ \label{eq:Um}
 U(\zeta)&= e^{\zeta \hat{a}_1\da \hat{a}_2 - \zeta^{\ast} \hat{a}_1 \hat{a}_2\da},
\end{align}  
 where $\xi:= s e^{i \phi_{B}}$ and $\zeta := m e^{i \phi_{A}}$, with  $s \geq 0$, $m \geq 0$ and $\phi_A, \phi_B \in \mathbb{R}$. In the following, we will consider the parameters of interest to  be the squeezing parameter $s$ and mode-mixing parameter $m$.  A relevant application for this interferometry scheme is considered in Section \ref{sec:GWDetector}.
 
  In order to formulate a simple expression for the QFI and sensitivity of this interferometry,  it is convenient to use the covariance matrix formalism.  This  uses a phase-space representation of a quantum state where a Gaussian state is fully defined by its displacement vector $\bs{d}$ and covariance matrix $\bs{\sigma}$.  In the real $q,p$ representation, these  are defined in the following way for a general system consisting of $n$ bosonic modes:\footnote{Note that several conventions are used for the definitions of $\bs{d}, \bs{\sigma}$ and $\hat{\bs{x}}$. See, e.g.\ \cite{CMF}, for an alternative convention to that presented here.}
\begin{align}
\bs{d} &:= \braket{\hat{\bs{x}}},\\
\bs{\sigma}_{ij} &:= \frac{1}{2} \braket{\{\hat{\bs{x}}_i, \hat{\bs{x}}_j\}} -  \braket{\hat{\bs{x}}_i} \braket{\hat{\bs{x}}_j},  
\end{align}

where $\hat{\bs{x}} := (\hat{x}_1, \hat{x}_2, \ldots  \hat{x}_{2n-1}, \hat{x}_{2n})^T$ and $\hat{x}_i$ are quadratures defined by:
\begin{align}
\hat{x}_{2i - 1} &:=  \hat{a}_i + \hat{a}_i\da,\\
\hat{x}_{2i} &:= i (\hat{a}_i\da - \hat{a}_i),
\end{align}
with $i \in \mathds{Z}^+$, and $\hat{a}_i$ and $\hat{a}_i\da$ the annihilation and creation operators.  For example, a two-mode  squeezed vacuum state is defined by $\bs{d} = \bs{0}$ and the covariance matrix:
\begin{align}
\bs{\sigma}  = 
\ba{cc} \cosh 2r ~\bs{1} & \sinh 2r ~\bs{R_{\vartheta_{sq}}} \\ \sinh 2r ~\bs{R_{\vartheta_{sq}}} & \cosh 2r ~\bs{1} \ea,
\end{align}
where:
\begin{align}
\bs{1} &:= \ba{cc} 1 & 0 \\ 0 & 1 \ea,\\
\bs{R_{\vartheta_{sq}}} &:= \ba{cc} \cos \vartheta_{sq} & \sin \vartheta_{sq} \\ \sin \vartheta_{sq} & -\cos \vartheta_{sq} \ea,
\end{align}
and $r$ and $\vartheta_{sq}$ are respectively the squeezing parameter and  phase. Unitary transformations $\bs{U}$ acting on density matrices now lead to symplectic matrices $\bs{S}$ acting on the displacement and covariance matrices through $\bs{d}' = \bs{S} \bs{d}$ and $\bs{\sigma} = \bs{S} \bs{\sigma} \bs{S}^T$ (see e.g.\ \cite{CMF}). 

The initial state of the pump mode is assumed to be a coherent state, and so the displacement and covariance matrices of the full  input state to the interferometer are:
\begin{align}
\bs{d}_0 &= \ba{c} 2 Re (\alpha) \\ 2 Im (\alpha) \\ 0 \\ 0 \\0 \\0\ea = \sqrt{\overline{N}} \ba{c} 2 \cos \vartheta_0 \\ 2 \sin \vartheta_0 \\ 0 \\ 0 \\0 \\0\ea,\\
\bs{\sigma}_0 &= \bs{1}_6,
\end{align}
where we have written $\alpha \equiv \sqrt{\overline{N}} e^{i \vartheta_0}$, with $\overline{N}$ the total particle number, and $\bs{1}_6 := \mathrm{diag}(1,1,1,1,1,1)$  the identity matrix of which the first two rows and columns are for the pump mode, the next two rows and columns are for one of the side modes, and the final rows and columns are for the other side mode. 

We next act on this state with a two-mode squeezing operation to parametrically populate the side modes. The state of the full system is now given by $\bs{S}_{s} \bs{d}_0$ and   $\bs{S}_{s} \bs{\sigma}_0 \bs{S}_{s} ^T$ where $\bs{S}_{s}$ is given by \eqref{eq:Ss} with  $r$ the squeezing parameter and  $\vartheta_{sq}$ the phase of the process. Here we have assumed that the pump is fairly undepleted by the squeezing operation and remains in a coherent state, but we take $\alpha \rightarrow \alpha_0$ after acting with $\bs{S}_s$, where  $|\alpha_0^2|:= |\alpha|^2 - 2 \sinh^2 r$, and $N_0 := |\alpha_0|^2$, $N := 2 \sinh^2 r$, so that particle number is conserved \cite{PumpedUpSU11}.  After the phonons are parametrically excited from the condensate, we apply a tritter to the three modes. The symplectic matrix for this operation $\bs{S}_{tr}$ is derived in Appendix \ref{app:TritterMatrix} and given by \eqref{eq:Str}, where $\theta$ is the angle of the tritter and $\vartheta$ is its phase. Following the tritter stage, we act on the side modes with the squeezing or mode-mixing operations given by \eqref{eq:Us} and \eqref{eq:Um}. The symplectic matrices for these operations are respectively given by \eqref{eq:Ssc} and \eqref{eq:Smc}. Subsequently, the beams are brought back together with another tritter and then an outcoupling process, which are both the reverse of the operations that were performed prior to the Gaussian unitary channel.  The state of the  full interferometer is then given by:
\begin{align} \label{eq:dEnd}
\bs{d} &= \bs{S} \bs{d}_0,\\ \label{eq:SigmaEnd}
\bs{\sigma} &= \bs{S}  \bs{\sigma}_0 \bs{S}^T,
\end{align}
where $\bs{S}:=\bs{S}_{-} \bs{S}_{\epsilon} \bs{S}_+$ with $\bs{S}_{-} := \bs{S}_s (-r) \bs{S}_t (-\theta)$, $\bs{S}_+ :=  \bs{S}_t (\theta) \bs{S}_s(r)$ and $\bs{S}_{\epsilon}$ being either the squeezing or mode-mixing channel for the side modes. At the end of the interferometer, we assume the conventional SU(1,1) number-sum measurement of particles in each mode \cite{SU11}. That is, the  the measured observable is: $\hat{S} = \hat{N} := \hat{a}^{\dagger}_1 \hat{a}_1 + \hat{a}^{\dagger}_2 \hat{a}_2$.

\subsection{Quantum fisher information}

Since it is independent of the particular measurement scheme used, when calculating the QFI we only need to consider the operations up to and including the Gaussian unitary channels i.e.\ the state of the relevant system is defined by $\bs{d} = \bs{S}_{\epsilon} \bs{S}_+ \bs{d}_0$ and $\bs{\sigma} = \bs{S}_{\epsilon} \bs{S}_+ \bs{d}_0  (\bs{S}_{\epsilon} \bs{S}_+)^T$.\footnote{To obtain the QFI for the conventional phase-shift channel, the $\bs{S}_{\epsilon}$ matrix would be that presented in \eqref{eq:Spc}. This then results in the same expressions derived in \cite{PumpedUpSU11} which were obtained using a Heisenberg picture with the annihilation and creation operators, in contrast to the Schr\"{o}dginer picture, phase-space method considered here.}  For Gaussian states, the QFI, $H_{\epsilon}$, can be obtained through \cite{QFIMatrices,QFIMatricesBraun}:
\begin{align} \label{eq:H}
H_{\epsilon} = \frac{1}{2} \frac{Tr[(\bs{\sigma}^{-1}_{\epsilon} \dot{\bs{\sigma}}_{\epsilon})^2 ]}{1 + \mu_{\epsilon}^2} + \Delta \bs{d}^{\prime T}_{\epsilon} \bs{\sigma}^{-1}_{\epsilon} \bs{d}^{\prime}_{\epsilon}  + 2 \frac{(\dot{\mu}_{\epsilon})^2}{1 - \mu_{\epsilon}^4} ,
\end{align}
where:
\begin{align}
\bs{d}_{\epsilon} &:= \bs{S}_{\epsilon} \bs{d},\\
\bs{\sigma}_{\epsilon} &:=  \bs{S}_{\epsilon}^T \bs{\sigma}\bs{S}_{\epsilon},\\
\dot{\bs{\sigma}_{\epsilon}} &:= \frac{d \bs{\sigma}_{\epsilon}}{d \epsilon },\\
\Delta \bs{d}^{\prime T}_{\epsilon} &:= \frac{d(   \bs{d}_{\epsilon + \xi} -  \bs{d}_{\epsilon})}{d \xi} \bigg\rvert_{\xi = 0},\\
\dot{\mu}_{\epsilon} &:= \frac{d \mu_{\epsilon}}{d \epsilon },
\end{align}
and $\mu_{\epsilon} := 1 / \sqrt{\det \bs{\sigma}_{\epsilon}}$ is the purity. Above we have taken $\epsilon$ as the parameter of interest, which is proportional to the  squeezing parameter $s$ of \eqref{eq:Us} for the squeezing operation, and $m$ of \eqref{eq:Um} for the mode-mixing operation. Precisely, we define $s=: \frac{1}{4} \epsilon B$ and $m=: \frac{1}{4} \epsilon A$ where $B$ and $A$ are  proportionality constants.

When the squeezing channel is chosen, the QFI is found to be:
	\begin{align} \label{eq:HSqFull}
	H &= \frac{1}{16} B^2 \Big[ 4 + \sin^2 (2 \theta) \sinh^2 r + 2 (1 + \cos^4 \theta) \eta_2(\vartheta_{sq}) \sinh^2 (2r) +|\alpha_0|^2 \Big(4 \sin^4 \theta + \eta_1(r) \sin^2 2 \theta \Big)\Big],
	\end{align}
where:
\begin{align}
\eta_1(r) &:= \sinh (2r) \cos (2 \vartheta - 2 \vartheta_0 - \vartheta_{sq} + 2 \phi_{B}) + \cosh (2r),\\
\eta_2(\vartheta_{sq}) &:= \sin^2 (\vartheta_{sq} - \phi_{B}).
\end{align}
When $\vartheta_{sq} = \phi_{B} +\pi/2$ and $\vartheta = \vartheta_0 - \phi_B/2 + \pi/4$, $H$ has three turning points at $\theta = 0$, $\theta = \pi/2$ and $\theta = \theta_t$ where $\theta_t$ is provided in \eqref{eq:thetat} and matches the analogous optimum angle found in \cite{PumpedUpSU11} for a phase-shift channel. For large $\overline{N}$, $\theta_t$ can be approximated as $\frac{1}{4} \pi + \csc^{-1} (N + \sqrt{N(N+2)}) / 2$. For $\theta = 0$, $\theta = \pi/2$ and $\theta = \theta_t$, $H$ is:
\begin{align} \label{eq:Hsq0}
H(\theta = 0) &= \frac{1}{4}B^2 \Big(1+ \sin^2 (\vartheta_{sq} - \phi_B) \sinh^2 (2r)\Big) \equiv \frac{1}{4} \Big(1+ \sin^2 (\vartheta_{sq} - \phi_B) N^2\Big),\\
H(\theta = \frac{\pi}{2}) &= \frac{1}{4}B^2 \Big(1+ N_0 + \frac{1}{2} \sin^2 (\vartheta_{sq} - \phi_B) N^2\Big),\\\label{eq:Hsqt}
H(\theta = \theta_t) &= \frac{1}{32}B^2  \overline{N}e^{2r} (1+ \coth r) + \mathcal{O}(\overline{N}^0) \\&\xrightarrow[r \gg 1]{}  \frac{1}{8}B^2 \overline{N} N,
\end{align}
where, for the $\theta = \theta_t$ turning point, we have assumed that $\overline{N} \gg 1$ and taken the optimum phase relation $\vartheta =  \vartheta_0 + \vartheta_{sq}/2 - 2 \phi_B$.	When $\theta = 0$ we recover standard SU(1,1) interferometry, and so \eqref{eq:Hsq0} is the QFI for an SU(1,1) interferometer with a squeezing channel. This type of  interferometer  derived when $\theta = 0$ could still be considered an SU(1,1) interferometer since 
the unitary representation of a squeezing channel is part of the SU(1,1) group and, although a phase-shift channel described by $\hat{S}$ is no longer present, $\hat{S}$ still forms the measurement process.

In general, if we assume $\overline{N} \gg 1$ in \eqref{eq:HSqFull}, then we obtain:
\begin{align} \label{eq:HSqLargeN}
H &= \frac{1}{4} B^2\overline{N}  \Big( \sin^4 \theta + \frac{1}{4}  \sin^2 (2 \theta) \eta_1 (r) \Big) + \mathcal{O} (\overline{N}^0)\\ \label{eq:HSqLargeNLim}
&\approx \frac{1}{8}B^2 \sin^2 (2 \theta) N \overline{N}, 
\end{align}
where in the last line we have assumed that $N \gg 2$ and taken  the optimum phase relation  $\vartheta =  \vartheta_0 + \vartheta_{sq}/2 - 2 \phi_B$. For the pump to remain relatively undepleted before the squeezing channel, we then have (see Appendix \ref{app:SmallTheta} for more detail):
\begin{align} \label{eq:Hsqsmalltheta}
H &\approx \frac{1}{4} B^2 \Big( 1 + N^2 + \theta^2 ( N_0 e^{2r} + N/2 - N^2) \Big)\\ \label{eq:pumpedQFIsq}
 &\approx \frac{1}{2} B^2 \theta^2 N_0 N ,
\end{align}
where we have again used $\vartheta =  \vartheta_0 + \vartheta_{sq}/2 - 2 \phi_B$, and further  assumed that $ r \gg 1$ and $\overline{N} \gg 1$ in the last line. Note that, as with pumped-up SU(1,1) with a phase-sift channel \cite{PumpedUpSU11}, the QFI for pumped-up SU(1,1) with a squeezing channel is never worse than a standard  SU(1,1) interferometer with a squeezing channel, and will likely be orders of magnitude larger in practical setups.  This is illustrated in Section  \ref{sec:GWDetector} using a particular practical application.

If  we instead use the mode-mixing channel, the QFI is given by:
\begin{align} \label{eq:HmFull}
H=\frac{1}{8} A^2 \Big[ &(1 + \cos^2 \theta) \sinh^2 (2r) + \sin^2  \theta ~\Phi_1(\theta,\phi_A) \Big( \sinh^2 (2r) - 2 \sinh^2 r\Big)\\&+2 |\alpha_0|^2 \sin^2 \theta \Big(\sin^2 \theta \sin^2 \phi + \Phi_1 (\theta,\phi_A) \eta_3 (r)\Big) \Big]
\end{align}
where:
\begin{align}
\Phi_1(\theta,\phi_A) &:= \sin ^2(\theta )  \sin^2( \phi_{A} )-1,\\
\eta_3(r) &:=  \sinh (2r) \cos (2 \vartheta - 2 \vartheta_0 + \vartheta_{sq} ) -  \cosh (2r).
\end{align}
Analogous expressions to \eqref{eq:Hsq0}-\eqref{eq:Hsqt} for the squeezing case can be obtained from \eqref{eq:HmFull} when $\theta = 0$, $\theta = \pi/2$ with $\phi_A = \pi/2$, and  $\theta = \pi/2$ with $\phi_A = 0$:
\begin{align} \label{eq:Hm0}
H(\theta = 0) &= \frac{1}{4}A^2 N^2,\\
H(\theta = \frac{\pi}{2}, \phi_A = \frac{\pi}{2}) &= \frac{1}{4}A^2 \Big( N_0 + \frac{1}{2} N^2\Big),\\
H(\theta = \frac{\pi}{2}, \phi_A =0 ) &= \frac{1}{4}A^2  (\overline{N}e^{2r} + N) + \mathcal{O}(\overline{N}^0) \\&\xrightarrow[r \gg 1]{}  \frac{1}{2} A^2 \overline{N} N,
\end{align}
where, for the last case we have assumed that $\overline{N} \gg 1$ and taken    $\vartheta =  \vartheta_0 - \vartheta_{sq}/2 +\pi/2$. As for the squeezing channel above, the $\theta = 0$ case \eqref{eq:Hm0} is the QFI for a conventional SU(1,1) interferometer with a mode-mixing channel. In contrast to the squeezing channel case, this type of interferometer derived when $\theta = 0$ would not be considered an SU(1,1) interferometer by the original definition \cite{SU11} since the unitary representation of the mode-mixing channel does not form part of the SU(1,1) group. Instead, such an interferometer would be described by a larger group, for example,  the unitary group associated with a double covering of Sp($4, \mathbb{R}$) \cite{Arvind1995}.

In general, if we assume $\overline{N} \gg 1$ in \eqref{eq:HmFull}, then we obtain:
\begin{align}\label{eq:HmLargeN}
H &= \frac{1}{4} A^2 \overline{N} \sin^2 \theta \Big( \sin^2 \theta \sin^2 \phi_A + \Phi_1(\theta,\phi_A) \eta_3(r)\Big)+ \mathcal{O} (\overline{N}^0)\\ \label{eq:HmLargeNLim}
&\approx \frac{1}{2 }A^2 \sin^2 \theta (1-\sin^2 \theta \sin^2 \phi_A)\overline{N} N , 
\end{align}
where in the last line we have assumed that $N \gg 1/ 2$ and again taken   $\vartheta =  \vartheta_0 - \vartheta_{sq}/2 +\pi/2$. For the pump to remain relatively
 undepleted before the mode-mixing channel, we then find:
\begin{align}  \label{eq:Hmsmalltheta}
H &\approx \frac{1}{4} A^2 \Big(  N^2 + \theta^2 ( N_0 e^{2r} + N/2 - N^2) \Big)\\ \label{eq:pumpedQFIm}
&\approx \frac{1}{2} A^2 \theta^2 N_0 N ,
\end{align}
where we have  assumed that $ r \gg 1$ in the last line. This is similar to the QFI for the squeezing channel \eqref{eq:Hsqsmalltheta}, just with $B$ replaced by $A$.  As in the squeezing case, the QFI for pumped-up SU(1,1) with a mode-mixing channel is never worse than the original SU(1,1) interferometry design  with a mode-mixing channel rather than a phase-shift channel, and will likely be orders of magnitude larger in practical setups, which we consider further  in Section  \ref{sec:GWDetector}.

\subsection{Sensitivity} \label{sec:Sensitivity}

In the previous section we calculated the QFI for pumped-up SU(1,1) interferometry with a mode-mixing and squeezing channel using the covariance matrix formalism.  We now consider the particular implementation of pumped-up SU(1,1) discussed in Section \ref{sec:GaussianChannels} where the measurement process is the sum of the number of particles in the side modes, i.e.\ the measured observable is $\hat{S} = \hat{N} := \hat{a}^{\dagger}_1 \hat{a}_1 + \hat{a}^{\dagger}_2 \hat{a}_2$. The square of the sensitivity of the interferometer is defined as (see e.g.\ \cite{QuantumLimitsOptics} for a derivation):
\begin{align} \label{eq:Sens}
\Delta^2 \epsilon := \frac{\mathrm{Var}(\hat{S})}{   (\partial_{\epsilon} \braket{\hat{S}})^2 }
\end{align}
where $\mathrm{Var}(\hat{S}) := \braket{\hat{S}^2} - \braket{\hat{S}}^2$. For Gaussian states, this is related to the Fisher information through \cite{GaussianActive}:
\begin{align}
F = F_0  + \frac{2 \Big(\partial_{\epsilon} \sqrt{\mathrm{Var}(\hat{S})}\Big)^2 }{\mathrm{Var}(\hat{S}) }
\end{align}
where:
\begin{align}
F_0 := \frac{1}{\Delta^2 \epsilon},
\end{align}
such that  $ F \geq F_0$.  Writing \eqref{eq:Sens} in the covariance matrix formalism in the $q,p$ basis, we find the simple expressions:\footnote{See Appendix \ref{app:Heterodyne} for the equivalent expressions for homodyne/heterodyne measurements.}
\begin{align} \label{eq:S}
\braket{\hat{S}} &= \frac{1}{4} [\mathrm{Tr}(\bs{\sigma}_s) + \bs{d}^T_s \bs{d}_s - 2n],\\ \label{eq:VarS}
\mathrm{Var}(\hat{S}) &= \frac{1}{8} [  \mathrm{Tr}(\bs{\sigma}^2_s) + 2 \bs{d}^T_s \bs{\sigma}_s \bs{d}_s - 2n] 
\end{align}
where $n$ is the number of modes, which is $2$ in this case, and $\bs{\sigma}_s$ and $\bs{d}_s$ are the covariance and displacement matrices of the side modes, which are generated from $\bs{d}$ and $\bs{\sigma}$ given in \eqref{eq:dEnd} and \eqref{eq:SigmaEnd} by simply removing the first two rows and columns.

Working with small $s$, if the squeezing channel is chosen, \eqref{eq:S}-\eqref{eq:VarS} become:
\begin{align}
\braket{\hat{S}} &= \frac{1}{4} s^2 \Big(|\alpha_0|^2 \sin ^2(2 \theta ) (\sinh (2 r) \cos (2 \vartheta -2 \vartheta_0-\vartheta_{sq}+2 \phi_{B} +\cosh (2 r))\\&\hspace{1cm}+4\left(1+ \cos ^4\theta \right) \left(\sinh ^2(2 r) \sin ^2(\vartheta_{sq}-\phi_{B} )+1\right)+\sin ^2(2 \theta ) \cosh ^2 r\Big),\\
\mathrm{Var}(\hat{S}) &= \frac{1}{4} s^2 \Big(|\alpha_0|^2 \sin ^2(2 \theta ) (\sinh (2 r) \cos (2 \vartheta -2 \vartheta_0-\vartheta_{sq}+2 \phi_{B} )+\cosh (2 r))\\&\hspace{1cm}+8\left(1+ \cos ^4 \theta \right) \left(\sinh ^2(2 r) \sin ^2(\vartheta_{sq}-\phi_{B} )+1\right)+\sin ^2(2 \theta ) \cosh ^2 r\Big),\\
\implies F_0 &=  \frac{1}{16} B^2 \frac{\Big((|\alpha_0|^2 \eta_1 + \cosh^2 r) \sin^2 (2 \theta) + \left(1+ \cos ^4 \theta \right) (\sinh^2 (2r) \eta_2 + 1) \Big)^2}{(|\alpha_0|^2 \eta_1 + \cosh^2 r) \sin^2 (2 \theta) + 2\left(1+ \cos ^4 \theta \right) (\sinh^2 (2r) \eta_2 + 1)}.
\end{align}
We now use particle conservation to write $|\alpha_0(r)|^2 = |\alpha|^2 - 2 \sinh^2 r$ and $|\alpha|^2 = \overline{N}$ so that $|\alpha(r)|^2 = 1/ \varepsilon  - 2 \sinh^2 r$, where $\varepsilon := 1 /\overline{N}$, and take $\overline{N} \gg 1$,  to find:
\begin{align}
F_0 = \frac{1}{16} B^2 \sin^2 (2 \theta) \eta_1(r) \overline{N} + \mathcal{O} (\overline{N}^{0}).
\end{align}
Note that this is contained in the  corresponding expression for the QFI  \eqref{eq:HSqLargeN} and agrees with \eqref{eq:HSqLargeNLim} in the same limits. For the pump to remain relatively undeleted before the squeezing channel, we then obtain   
\begin{align} \label{eq:F0Squeezed}
F_0 \approx \frac{1}{2} B^2 \theta^2 N_0 N,
\end{align}
when  $r \gg 1$, and  $2 \vartheta = 2 \vartheta_0 + \vartheta_{sq} - 2 \phi_{B}$, which agrees with the corresponding QFI expression \eqref{eq:pumpedQFIsq}. The  number-sum measurement is, therefore, an optimum measurement scheme in these limits.

On the other hand, if the mode-mixing channel is chosen, then we have:
\begin{align}
\braket{\hat{S}} &=   m^2 \Big( \sin^2 \theta \left( \Phi_1 |\alpha_0| ^2  \eta_3(r) -  \Phi_2 \sinh ^2 r +(\Phi_1-1) \sinh ^2(2 r)  \right)+2 \sinh ^2(2 r)\Big),\\
\mathrm{Var}(\hat{S}) &= m^2 \Big(\Phi_1  \sin ^2\theta \left(|\alpha_0|^2 \eta_3(r)-\sinh ^2 r+ 2  \sinh ^2(2 r)\right)+2 (1 + \cos^2 \theta) \sinh ^2(2 r)\Big),
\end{align}
where
\begin{align}
\Phi_2 &:= \sin ^2(\theta )  \cos^2( \phi_{A} )-1.
\end{align}
Taking $\overline{N} \gg 1$, we find:
\begin{align}
F_0 &= \frac{1}{4} A^2 \sin^2 \theta~ \Phi_1 (\theta,\phi_A) \eta_3(r) \overline{N} +  \mathcal{O} (\overline{N}^{0}).
\end{align}
Note that, as with the squeezing channel considered above, this is contained in the  corresponding expression for the mode-mixing  QFI  \eqref{eq:HmLargeN} and agrees with \eqref{eq:HmLargeNLim} in the same limits. For the pump to remain relatively undeleted before the mode-mixing channel, we then obtain:   
\begin{align} \label{eq:F0ModeMixed}
F_0 \approx \frac{1}{2} A^2 \theta^2 N_0 N,
\end{align}
with $2 \vartheta = 2 \vartheta_0 - \vartheta_{sq}$ and $r \gg 1$. This agrees with the corresponding QFI expression \eqref{eq:pumpedQFIm}. The  number-sum measurement is, therefore, also an optimum measurement scheme for the mode-mixing case in these limits.

\section{Application: detecting gravitational waves with phonons of a BEC} \label{sec:GWDetector}

We now apply our considered interferometry schemes to Bose-Einstein condensates and, in particular, a GW detector based on phonons of a BEC that resonate with a GW. This detector was recently  proposed in \cite{GWDetectorFirst} and  was suggested to look for  GWs from persistent sources that have frequencies slightly higher than the GWs suited to LIGO (around $ 1\,\mathrm{kHz}-100\,\mathrm{kHz}$). Persistent sources include, for example, pulsars and spinning neutron stars (that are non-axisymmetric), and, in contrast to the GWs from mergers seen by LIGO and VIRGO,  GWs from persistent sources have not been yet been observed. The detector uses the collective quantum excitations (phonons) of a BEC, which are resonantly squeezed or mode-mixed by a GW in a scheme that resembles a quantum version of resonant mass detectors \cite{WeberBar},  but with the GW parametrically, rather than directly, driving the quanta of sound waves (see e.g.\ \cite{ReviewPaper} for a discussion on the similarities and differences of these two types of detectors). The detector also utilizes quantum metrology. In particular, a two-mode squeezed state of phonons is prepared that acts as a probe state in a quantum metrology scheme for GW detection, and Heisenberg scaling in the number of initial phonons is observed.\footnote{Rather than considering phonons of a BEC in an effectively rigid trap, we could also imagine a squeezed state of light in an optical cavity (optical resonator), which will be the concern of future work (see also, e.g.\ \cite{braginskii1974electromagnetic,ReviewPaper}, for alternative schemes using optical resonators). The GWs that would resonate with the light field could then be much higher than those considered here, with possible sources including, for example, exploding Planck stars \cite{PlanckStars}.}  

Since the GW acts as a squeezing or mode-mixing channel on the phonons (at least to second order in the GW strain) we can, in principle, apply the interferometry setup considered in the previous sections to this detector. This new scheme can  be thought of as mixing the interferometry setup of the highly successful laser GW detectors, such as LIGO and VIRGO, with the resonance process most associated with resonant-mass detectors, which were the first GW detectors \cite{WeberBar}. However,  parametric resonance is utilized rather than the direct resonance process found  in traditional resonant-mass detectors, which would represent a coherent and so classical channel. 

To apply the considered interferometry scheme,  we take the condensate to be the pump mode (since its state can be approximated by a highly-populated coherent state), and the two phonon modes to be the side modes.  The scheme requires parametrically amplifying the phonon modes from the condensate during the initial active element stage, and mixing the condensate with the phonon modes during the tritter stage. The former could be achieved, for example, by modifying the boundary conditions of the phonons in an analogue to the  dynamical Casimir effect \cite{DCETheory,DCEWestbrook,Accelerometer,GWDetectorFirst,SchmiedmayerDCE},  introducing an additional time-dependent potential \cite{MassBEC} (also see  Appendix \ref{app:implementation}), applying a laser to the BEC \cite{PhysRevLett.115.060401,PhysRevA.93.023610,Steinhauer2016}, or using other methods such as enhancing Beliaev damping \cite{BeliaevEntanglement}. For the tritter, phonons have already been beam split with a condensate in a process resembling heterodyne detection by turning off the trapping potential \cite{PhononEvap,HeterodyneBECs}. However, here we propose implementing the mixing at the initial stage of an interferometer  rather than just for a quantum measurement process (see  Appendix \ref{app:implementation} for a possible implementation that again involves applying an additional, oscillating potential to the BEC). This would mean that all the active and passive elements of the interferometer would consist of operations applied to a single trap that contains the BEC. That is, the elements would not be spatially separated as in a traditional optical interferometer.

\subsection{Quantum Fisher Information}

In \cite{GWDetectorFirst}, a two-mode squeezed state of phonons is prepared and then a GW is found to effectively acts as a squeezing or mode-mixing channel on the phonons. The QFI for the resonant squeezing channel can be shown to  be (see Appendix \ref{app:GWDetector} for more detail):
\begin{align} 
H &= \frac{1}{4} B^2 [ 1 + \sin^2 (\vartheta_{sq}- \phi_{B}) \sinh^2 2r]\\ \label{eq:QFI}
&= \frac{1}{4} B^2 [ 1 + N^2_P] ~~\mathrm{when}~~  \vartheta_{sq} = \phi_{B} + \frac{\pi}{2},
\end{align}
where $N_P = 2 \sinh^2 r$ is the number of initially squeezed phonons and $B$ is defined in \eqref{eq:BGW} with the angular frequencies  $\omega_m,\omega_n$ of two phonon modes satisfying  $\Omega = \omega_m + \omega_n$ where $\Omega$ is the GW angular frequency. The QFI, therefore, has Heisenberg scaling, which is the optimum scaling achievable with a quadratic Hamiltonian. In general, the sensitivity of the detector is bounded by the QCRB:\footnote{Here we are just concentrating on the sensitivity to the GW strain as defined by the  QCRB rather than, for example, the spectral strain sensitivity as often defined for GW detectors (see e.g.\ \cite{maggiore2008gravitational}).}  
\begin{align} \label{eq:QCRB}
\Delta \epsilon &\geq \frac{1}{\sqrt{M H}}\\
&\geq \frac{1}{\sqrt{ \frac{1}{4} B^2 N_P^2 N_d \tau/t}},
\end{align}
where $N_d$ is the number of detectors, $\tau$ is the integration time, and we have assumed that $N_P \gg 1$ and that there is no delay in repeating the experiments.

If we instead use a pumped-up SU(1,1) interferometry scheme as outlined above, then the GW would be considered to act on the phonons once they had been sent through a tritter with the condensate. The QFI in this new case would be given by \eqref{eq:pumpedQFIsq} with $B$ given by \eqref{eq:BGW}. Since $N_0$ is always much greater than $N$ for the description of the BEC used here, and taking $\theta = 0$ in \eqref{eq:Hsqsmalltheta}  recovers the QFI of the original scheme, then the QFI is never worse in the SU(1,1) case. Furthermore, it is likely to be experimentally challenging to create large phonon states in a  squeezed vacuum state, and the SU(1,1) scheme utilizes the large number of atoms in the condensate to compensate for this.  Taking, for example, $N_0 = 10^6$ and $r = 4.2$ in \eqref{eq:QFI}, then the same QFI can be generated using $r=2$ (and $\theta^2 \approx 0.094$ so that our description of the BEC remains valid during the GW channel) with the QFI for the pumped-up scheme \eqref{eq:pumpedQFIsq}, and keeping $r=4.2$ (with $\theta^2 \approx 0.092$) leads to two orders of magnitude improvement over that obtained from  \eqref{eq:QFI}.\footnote{This could potentially be improved further if the condensate could also be squeezed, perhaps allowing for a $N_0^2$ Heisenberg-like scaling. However, this would likely require a modified description for the BEC state that the GW interacts with than previously considered in \cite{GWDetectorFirst}.} For the mode-mixing channel of the detector, on the other hand, the QFI would be given by \eqref{eq:pumpedQFIm} in the pumped-up SU(1,1) scheme, with $A$ given by \eqref{eq:AGW}. 

\subsection{Sensitivity}

In general, the sensitivity of the detector in the pumped-up SU(1,1) scheme is  bounded by \eqref{eq:QCRB} but now with $H$ given respectively by \eqref{eq:pumpedQFIsq} and \eqref{eq:pumpedQFIm} for the squeezing and mode-mixing channels. One possible implementation of the original scheme proposed in \cite{GWDetectorFirst} is  to prepare an experiment similar to that in \cite{DCEWestbrook} where squeezed phonons could be created by oscillating the trapping potential, and then phonon evaporation and singe-atom detectors are used to measure the phononic state after the GW.  The classical Fisher information for such an experiment was calculated in \cite{GWDetectorThermal} and found to well-approximate the QFI given by \eqref{eq:QFI}. On the other hand, with the pumped-up SU(1,1) scheme using a number-sum measurement, the sensitivity would be given by \eqref{eq:F0Squeezed} and \eqref{eq:F0ModeMixed} for the different quantum channels, with the $A$ and $B$ definitions given above. As shown in Section \ref{sec:Sensitivity}, these sensitivities approximate that which would be obtained from the QFIs when $N_0 \gg N$, which is the case considered here.

\section{Summary} \label{sec:Summary}

We have introduced an active interferometric scheme based on the pumped-up SU(1,1) interferometer  \cite{PumpedUpSU11}, but with mode-mixing and squeezing channels rather than a conventional phase-shift channel. The QFI and, assuming a number-sum measurement, sensitivity of the interferometer can achieve similar scaling to that observed in the original pumped-up scheme \cite{SU11} and should, in practice, provide orders of magnitude improvement over the original SU(1,1) interferometer with the phase-shift channel replaced by a Gaussian channel. We have calculated these quantities using the covariance matrix formalism, finding simple and convenient expressions  for the sensitivity (see \eqref{eq:S}-\eqref{eq:VarS}).  We have also applied this interferometer setup to a GW detector based on phonons of a BEC since here the GW is expected to essentially act as a mode-mixing or squeezing channel on the phonons in a resonant process (see Appendix \ref{app:GWDetector}).  In practice, this can potentially improve the sensitivity of the detector compared to its original formulation \cite{GWDetectorFirst} by several orders of magnitude.

\newpage
\appendix
\section{Symplectic matrices of  interferometry operations}

Here we provide the symplectic matrices, in the real $q,p$ representation, for the various processes involved in our considered active interferometry schemes, as illustrated in Figure \ref{fig:SU11} and discussed in Section \ref{sec:GaussianChannels}. The first stage of the interferometer is   the  two-mode squeezing operation that parametrically populates the side modes, which has the following  symplectic matrix (see e.g.\ \cite{CMF}): 
\begin{align} \label{eq:Ss}
\bs{S}_{s} &= \ba{cccccc} 1 & 0 & 0& 0 & 0 & 0 \\
0 & 1 & 0& 0 & 0 & 0 \\
0 & 0 & \cosh r & 0  & \sinh r \cos \vartheta_{sq} & \sinh r \sin \vartheta_{sq} \\
0 & 0 & 0 & \cosh r & \sinh r \sin \vartheta_{sq} & - \sinh r \cos \vartheta_{sq} \\
0 & 0 & \sinh r \cos \vartheta_{sq} & \sinh r \sin \vartheta_{sq} & \cosh r & 0 \\
0 & 0 & \sinh r \sin \vartheta_{sq} & - \sinh r \cos \vartheta_{sq} & 0 & \cosh r \ea,
\end{align}
where $r$ is the squeezing parameter and $\vartheta_{sq}$ is the squeezing phase. The same convention is used as in Section \ref{sec:GaussianChannels} such that the first two columns and rows are for the pump, the next two column and rows are for one of the side modes, and the last two columns and rows are for the other side mode.

The next stage is  a tritter between the side-modes and  the pump, which has the following symplectic matrix (see Appendix \ref{app:TritterMatrix} for its derivation):
\begin{align} \label{eq:Str}
\bs{S}_{tr} &=\ba{cccccc} \cos \theta & 0 & \frac{1}{\sqrt{2}} \sin \theta \sin \vartheta & \frac{1}{\sqrt{2}} \sin \theta \cos \vartheta & \frac{1}{\sqrt{2}} \sin \theta \sin \vartheta  & \frac{1}{\sqrt{2}} \sin \theta \cos \vartheta \\
0 & \cos \theta & - \frac{1}{\sqrt{2}} \sin \theta \cos \vartheta  & \frac{1}{\sqrt{2}} \sin \theta \sin \vartheta  & - \frac{1}{\sqrt{2}} \sin \theta \cos \vartheta & \frac{1}{\sqrt{2}} \sin \theta \sin \vartheta  \\
- \frac{1}{\sqrt{2}} \sin \theta \sin \vartheta  & \frac{1}{\sqrt{2}} \sin \theta \cos \vartheta & \cos^2 (\frac{\theta}{2}) & 0 & \frac{1}{2} (-1 + \cos \theta) & 0 \\
-\frac{1}{\sqrt{2}} \sin \theta \cos \vartheta & \frac{1}{\sqrt{2}} \sin \theta \sin \vartheta  & 0 & \cos^2 (\frac{\theta}{2}) & 0 & \frac{1}{2} (-1 + \cos \theta) \\ -\frac{1}{\sqrt{2}} \sin \theta \sin \vartheta   & \frac{1}{\sqrt{2}} \sin \theta \cos \vartheta & \frac{1}{2} (-1 + \cos \theta) & 0 & \cos^2 (\frac{\theta}{2}) & 0\\
-\frac{1}{\sqrt{2}} \sin \theta \cos \vartheta & -\frac{1}{\sqrt{2}} \sin \theta \sin \vartheta & 0 & \frac{1}{2} (-1 + \cos \theta) & 0 & \cos^2 (\frac{\theta}{2}) \ea,
\end{align}
where $\theta$ is the angle of the tritter and $\vartheta$ is its phase (see Appendix \ref{app:TritterMatrix} for their definitions in terms of the Hamiltonian of the tritter).

Following the tritter, there is the squeezing or mode-mixing channel, which are defined by the unitary transformations \eqref{eq:Us}-\eqref{eq:Um}. The symplectic matrices for these are (see e.g.\ \cite{CMF}):
\begin{align} \label{eq:Ssc}
\bs{S}_{sc} &= \ba{ccc} \bs{1} & \bs{0} & \bs{0} \\
\bs{0} &  \bs{1} \cosh s &   \bs{R_{\phi_{B}}} \sinh s \\
\bs{0} &  \bs{R_{\phi_{B}}} \sinh s &   \bs{1} \cosh s \ea ,
\end{align}
 and 
\begin{align} \label{eq:Smc}
\bs{S}_{mc} &= \ba{ccc} \bs{1} & \bs{0} & \bs{0} \\
\bs{0} & \bs{1} \cos  m  &   \bs{R_{ \phi_{A}}} \sin m \\
\bs{0} & -\bs{R^T_{ \phi_{A}}}\sin m  &  \bs{1} \cos  m \ea,
\end{align}
where:
\begin{align}
\bs{R_{ \phi_{B}}} &:= \ba{cc} \cos \phi_{ {B}} & \sin \phi_{ {B} } \\ \sin \phi_{ {B}} &  -\cos \phi_{ {B}}\ea,\\
\bs{R_{ \phi_{A}}} &:= \ba{cc} \cos \phi_{ {A}} & \sin \phi_{ {A} } \\ -\sin \phi_{ {A}} &  \cos \phi_{ {A}}\ea.
\end{align}
In contrast, the symplectic matrix for a unitary phase evolution  $\hat{U}(\phi) = \exp (-i \phi \hat{N}/2)$ would be the following (see e.g.\ \cite{CMF}):
\begin{align} \label{eq:Spc}
\bs{S}_{pc} = \ba{ccccccc} 1 & 0 & 0& 0& 0& 0 \\ 0 & 1 & 0& 0& 0& 0 \\ 0 & 0 & \cos \frac{\phi}{2} & \sin \frac{\phi}{2} & 0 & 0\\
0 & 0& -\sin \frac{\phi}{2} & \cos \frac{\phi}{2} & 0 & 0 \\
0 & 0 & 0 & 0 & \cos \frac{\phi}{2}& \sin \frac{\phi}{2} \\
0 & 0 & 0 & 0 & -\sin \frac{\phi}{2}  & \cos \frac{\phi}{2} \ea.
\end{align}

\section{Derivation of symplectic matrix of  tritter}  \label{app:TritterMatrix}

The tritter used in the pumped-up SU(1,1) interferometry scheme  is generated by the following Hamiltonian \cite{PumpedUpSU11}:
\begin{align} \label{eq:Htritter}
H_{tr} = \frac{\hbar G}{\sqrt{2}} \Big[ e^{i \vartheta} \hat{a}_0^{\dagger} (\hat{a}_1 + \hat{a}_2) + e^{-i \vartheta} \hat{a}_0 (\hat{a}^{\dagger}_1 + \hat{a}^{\dagger}_2)  \Big],
\end{align}
which, in the Heisenberg picture, results in:
\begin{align}
\hat{a}_{1,2} (\theta) &= \hat{a}_{1,2} \cos^2 (\theta/2) - \hat{a}_{1,2} \sin^2 (\theta / 2) - \frac{i e^{- i\vartheta}}{\sqrt{2}} \hat{a}_0 \sin \theta,\\
\hat{a}_0 (\theta) &= \hat{a}_0 \cos \theta - \frac{ i e^{i \vartheta}}{\sqrt{2}} (\hat{a}_1 + \hat{a}_2) \sin \theta .
\end{align}
We can write this as:
\begin{align}
\bs{a}(\theta) :=\bs{A} \bs{a}
\end{align}
where:
\begin{align}
\bs{a} :=  \ba{c} \hat{a}_0 \\ \hat{a}^{\dagger}_0 \\ \hat{a}_1\\ \hat{a}_1^{\dagger}  \\ \hat{a}_2 \\ \hat{a}_2^{\dagger} \ea,
\end{align}
\begin{align}
\bs{A} := \ba{cccccc} \cos \theta & 0&  -\frac{i e^{i\vartheta}}{\sqrt{2}}  \sin \theta & 0 & -\frac{i e^{i\vartheta}}{\sqrt{2}}  \sin \theta & 0 \\
0 & \cos \theta&  0& \frac{i e^{-i\vartheta}}{\sqrt{2}}  \sin \theta & 0 & \frac{i e^{-i\vartheta}}{\sqrt{2}}  \sin \theta  \\
-\frac{i e^{-i\vartheta}}{\sqrt{2}} & 0 & \cos^2 (\theta/2) & 0 &- \sin^2 (\theta/2) &0\\
0 & -\frac{i e^{i\vartheta}}{\sqrt{2}} & 0 & \cos^2 (\theta/2) & 0 &- \sin^2 (\theta/2)\\
-\frac{i e^{-i\vartheta}}{\sqrt{2}} & 0 & 
- \sin^2 (\theta/2) & 0 &\cos^2 (\theta/2) &0\\
0 & -\frac{i e^{i\vartheta}}{\sqrt{2}} & 0 & 
\cos^2 (\theta/2) & 0  &\cos^2 (\theta/2)
\ea  .
\end{align}
We now move to the real symplectic $q,p$ representation. In this case we have:
\begin{align}
\bs{q}  = \bs{Q} \bs{a},
\end{align}
where:
\begin{align}
\bs{q} := \ba{c} q_0 \\ p_0 \\ q_1 \\ p_1 \\ q_2 \\ q_2 \ea, 
\end{align}
and:
\begin{align}
\bs{Q} := \ba{cccccc} 1 & 1 & 0 & 0 & 0 &0 \\
-i & i & 0 & 0 & 0 &0 \\
0 & 0 & 1 & 1 & 0 &0 \\
0 & 0 & -i & i & 0 &0\\
0 & 0 & 0 & 0 & 1 &1\\
0 & 0 & 0 & 0 & -i &i
\ea.
\end{align}
The symplectic representation of the tritter transformation is then:
\begin{align}
\bs{q}(\theta) = \bs{Q} \bs{a} (\theta) = \bs{Q} \bs{A} \bs{a} = \bs{Q} \bs{A} \bs{Q}^{-1}  \bs{q} := \bs{S}_{tr} \bs{q},
\end{align}
where:
\begin{align}
\bs{S}_{tr} &= \bs{Q} \bs{A} \bs{Q}^{-1}\\
&=\bsm \cos \theta & 0 & \frac{1}{\sqrt{2}} \sin \theta \sin \vartheta & \frac{1}{\sqrt{2}} \sin \theta \cos \vartheta & \frac{1}{\sqrt{2}} \sin \theta \sin \vartheta  & \frac{1}{\sqrt{2}} \sin \theta \cos \vartheta \\
0 & \cos \theta & - \frac{1}{\sqrt{2}} \sin \theta \cos \vartheta  & \frac{1}{\sqrt{2}} \sin \theta \sin \vartheta  & - \frac{1}{\sqrt{2}} \sin \theta \cos \vartheta & \frac{1}{\sqrt{2}} \sin \theta \sin \vartheta  \\
- \frac{1}{\sqrt{2}} \sin \theta \sin \vartheta  & \frac{1}{\sqrt{2}} \sin \theta \cos \vartheta & \cos^2 (\frac{\theta}{2}) & 0 & \frac{1}{2} (-1 + \cos \theta) & 0 \\
-\frac{1}{\sqrt{2}} \sin \theta \cos \vartheta & \frac{1}{\sqrt{2}} \sin \theta \sin \vartheta  & 0 & \cos^2 (\frac{\theta}{2}) & 0 & \frac{1}{2} (-1 + \cos \theta) \\ -\frac{1}{\sqrt{2}} \sin \theta \sin \vartheta   & \frac{1}{\sqrt{2}} \sin \theta \cos \vartheta & \frac{1}{2} (-1 + \cos \theta) & 0 & \cos^2 (\frac{\theta}{2}) & 0\\
-\frac{1}{\sqrt{2}} \sin \theta \cos \vartheta & -\frac{1}{\sqrt{2}} \sin \theta \sin \vartheta & 0 & \frac{1}{2} (-1 + \cos \theta) & 0 & \cos^2 (\frac{\theta}{2}) \esm .
\end{align}
Note that with a conventional two-way beam-splitter, $\theta = \pi/2$ would swap the modes. However, for the above tritter, $\theta = \pi/2$ would not completely swap the side modes and pump modes. This is responsible for $N$ appearing in the QFI expressions in Section \ref{sec:GaussianChannels}, even when $\theta = \pi/2$.

\section{Quantum Fisher Information}

The QFI for the SU(1,1) scheme presented in Section \ref{sec:GaussianChannels} when there is the squeezing channel is given by \eqref{eq:HSqFull}. Taking  $\vartheta_{sq} = \phi_B + \pi/2$ and $\vartheta = \vartheta_0 - \phi_B/2 + \pi/4$, the QFI $H$ has three turning points: $\theta = 0$, $\theta = \pi/2$ and $\theta = \theta_t$ where $\theta_t = \cos^{-1} (z_t)/2$ with:
\begin{align} \label{eq:thetat}
\theta_t &:= \cos^{-1} (z_t)/2,\\
z_t &:= \frac{\csch^2 r (\sinh(2r)^2 - 2|\alpha_0|^2)}{4 |\alpha_0|^2(1+ \coth r) - 2 \cosh(2r)} \equiv \frac{N (N+4) - 2 \overline{N}}{N (2 \overline{N} - 3 N - 1) + 2(\overline{N} - N) \sqrt{N(N+2)}}.
\end{align}
 The angle $\theta_t$ matches that  found in \cite{PumpedUpSU11} for a phase-shift channel. When $\overline{N}$ is large it can be approximated by  \cite{PumpedUpSU11}:
\begin{align}
\theta_t \approx \frac{1}{4} \pi + \frac{1}{2} \csc^{-1} (N + \sqrt{N(N+2)}).
\end{align}

\section{Full undepleted pump regime} \label{app:SmallTheta}

In Section \ref{sec:GaussianChannels}, we assumed that the pump is relatively undepleted after the first active element (see also \cite{PumpedUpSU11}). If we want to further assume that the pump is also relatively undepleted after the tritter stage, then $\theta$ cannot be too large.  After the tritter stage, in general, the number of particles in the pump and side modes is the following:
\begin{align}
N_0(\theta) &= N_0 \cos^2 \theta + \frac{1}{2} N \sin^2 \theta,\\
N(\theta) &= N_0 \sin^2 \theta + \frac{1}{2} N (1 + \cos^2 \theta ).
\end{align}
Let us require that $N = \gamma N_0$ and $N(\theta) = \delta N_0 (\theta)$  where $\gamma \ll 1$, $\delta \ll 1$ and $\delta \geq \alpha$. Then $\theta$ must satisfy:
\begin{align}
\theta \leq \frac{1}{2}  \arccos \Big(\frac{\delta \gamma + 2 \delta - 3 \gamma - 2}{\delta \gamma - 2 \delta + \gamma - 2}\Big).
\end{align} 
For example, taking $\delta = 0.1$,  we obtain:
\begin{align}
\theta \leq \frac{1}{2}  \arccos \Big(\frac{18 + 29 \gamma}{22 - 11 \gamma}\Big),
\end{align} 
which, in the limit $\gamma \rightarrow 0$, gives¬ $\theta^2 \approx 0.0938$, and we note that $\theta^2 / \sin^2 \theta \approx 1.03$.

\section{Experimental implementation of parametric amplifier and tritter in BECs} \label{app:implementation}

The quantum field Hamiltonian for a rarefied, interacting, non-relativistic Bose gas   can be well-approximated by (see, e.g.\ \cite{PitaevskiiBook})
\begin{align} \label{eq:fullH}
\hat{H} =  &\int  d\boldsymbol{r} \hat{\Psi}^{\dagger} \Big[-\frac{\hbar^2}{2m} \nabla^2 + \mathcal{V}(\boldsymbol{r}) \Big]\hat{\Psi} 
+ \frac{1}{2} g \int  d\boldsymbol{r} \hat{\Psi}^{\dagger} \hat{\Psi}^{\dagger} \hat{\Psi} \hat{\Psi},
\end{align}
where $\hat{\Psi}$ is the atomic field operator; $\mathcal{V}(\boldsymbol{r})$ is the trapping potential; and $g = 4 \pi \hbar^2 a / m$   is the coupling strength for a two-body contact  potential, with $a$  the s-wave scattering length and $m$  the atomic mass \cite{PitaevskiiBook,LandauLifshitzQM}.

Working in the Heisenberg  picture, the field $\hat{\Psi}$ can be decomposed into:
\begin{align} \label{eq:PsiExpand}
\hat{\Psi} (\boldsymbol{r},t) &=  [\hat{\psi}_0(\bs{r}) + \hat{\psi}(\bs{r},t) ] e^{- i \mu(t) / \hbar},
\end{align}
where:
\begin{align}
\hat{\psi}_0(\bs{r}) &:= \phi_0(\bs{r}) \hat{a}_0,\\
\hat{\psi}(\bs{r},t) &:= \sum_{n \neq 0} \phi_{n} (\bs{r}) \hat{a}_{n}(t),
\end{align}
with $\hat{a}_0$ the annihilation operator for the ground state,  $\hat{a}_n$ the annihilation operator for the $n$th excited state, and $\mu(t) = \mu t$, with $\mu$ the chemical potential. We now apply the Bogoliubov approximation where we assume that the ground state is macroscopically occupied such that removing one atom has little effect on the behaviour of the system,  so $\hat{a}_0 \ket{N_0} = \sqrt{N_0} \ket{N_0-1} \approx \sqrt{N_0} \ket{N_0}$. The ground state is then effectively in a large coherent state (i.e.\ approximately in a classical state):  $\hat{a}_0 \ket{\psi_0} = \sqrt{N_0} \ket{\psi_0}$ and $\hat{a}\da_0 \ket{\psi_0} \approx \sqrt{N_0} \ket{\psi_0}$, so that we can drop the hat from $\hat{a}_0$ and $\hat{a}\da_0$. We also apply a Bogoliubov transformation to $\hat{a}_n$ such that we can write the excited field $\hat{\psi}$ as:
\begin{align} \label{eq:BogoTransP} 
\hat{\psi}(\bs{r},t) = \sum_n [ u_n (\bs{r}) \hat{b}_n e^{- i\omega_n t} + v^{\ast}_n (\bs{r}) \hat{b}\da_n e^{i \omega_n t} ].
\end{align}
Then, plugging \eqref{eq:PsiExpand} into \eqref{eq:fullH}, using the Bogoliubov approximation and transformation  \eqref{eq:BogoTransP}, and dropping terms trilinear and quartic in $\hat{b}_n,\hat{b}\da_n$ (since they have fewer factors of $\sqrt{N_0} \gg 1$), the Hamiltonian is diagonalized as (see e.g.\ \cite{PitaevskiiBook}):
\begin{align}
:\hat{H}: ~= \sum_n \hbar \omega_n \hat{b}\da_n \hat{b}_n
\end{align}
where $: :$ refers to normal ordering; $u_n,v_n,$ and $\omega_n$ satisfy the Bogoliubov-de-Gennes equations:
\begin{align}
\hbar \omega_n u_n(\bs{r}) &= \Big[ -\frac{\hbar^2}{2m} \nabla^2 + \mathcal{V} (\bs{r}) - \mu + 2 g N_0 |\phi_0|^2 \Big] u_n (\bs{r}) +  g N_0 \phi_0^2 v_n (\bs{r})\\  
-\hbar \omega_n v_n(\bs{r}) &= \Big[ -\frac{\hbar^2}{2m} \nabla^2 + \mathcal{V} (\bs{r}) - \mu + 2 g N_0 |\phi_0|^2 \Big] v_n (\bs{r}) + g N_0 \phi^{\ast 2}_0 u_n (\bs{r}),
\end{align}
$u_n$ and $v_n$ are orthonormal \cite{PitaevskiiBook}:
\begin{align}
\int d \bs{r} [ u_n^{\ast} (\bs{r}) u_m (\bs{r}) - v^{\ast}_n (\bs{r}) v_m (\bs{r}) ] = \delta_{nm},
\end{align}
and $\phi_0$ satisfies the time-independent Gross-Pitaevskii equation:
\begin{align}
\Big[ -\frac{\hbar^2}{2m} \nabla^2 + \mathcal{V} (\bs{r}) +  g N_0 |\phi_0|^2 \Big]\phi_0  = \mu \phi_0,
\end{align}
such that $\phi_0(t) := \phi_0 e^{-i \mu t / \hbar}$ satisfies the time-dependent version.

We now apply a small time-dependent potential $\epsilon \mathcal{V}_{\epsilon} (\bs{r},t)$ to the BEC where $\epsilon \ll 1$. This introduces a term $\epsilon \mathcal{V}_{\epsilon} \hat{\Psi}\da \hat{\Psi}$ to \eqref{eq:fullH}, which, after applying \eqref{eq:PsiExpand} and \eqref{eq:BogoTransP}, provides an interaction Hamiltonian (see Appendix C of \cite{MassBEC} for a detailed derivation using the grand canonical Hamiltonian):
\begin{align} \label{eq:Hint}
\hat{H}_I (t) = \epsilon \int d \bs{r} \mathcal{V}_{\epsilon} (\bs{r},t) \Big[ |a_0|^2 |\phi_0|^2  &+ |a_0| \sum_{n} \Big(   \hat{b}_n e^{- i \vartheta (\bs{r}) } e^{-i\omega_n t}  +   \hat{b}\da_n e^{i \vartheta (\bs{r}) }  e^{i\omega_n t}\Big)\\
&+\sum_{n,m} [u_n^{\ast} (\bs{r}) u_m (\bs{r}) \hat{b}_n\da \hat{b}_m e^{i(\omega_m - \omega_n) t} + v_n (\bs{r})  v_m^{\ast}  (\bs{r}) \hat{b}_n \hat{b}_m\da e^{-i(\omega_m - \omega_n) t}] \\
&+\sum_{n,m} [u_n^{\ast}  (\bs{r}) v^{\ast}_m  (\bs{r}) \hat{b}_n\da \hat{b}\da_m e^{i(\omega_m + \omega_n) t} + u_n  (\bs{r}) v_m  (\bs{r}) \hat{b}_n \hat{b}_m e^{-i(\omega_m + \omega_n) t}]\Big],
\end{align}
where $\exp (i \vartheta(\bs{r})):= \phi_0 (\bs{r}) u^{\ast}_n (\bs{r}) + \phi^{\ast}_0 (\bs{r}) v^{\ast}_n (\bs{r})$.  We now investigate how this interaction Hamiltonian can be used to generate a two-mode squeezed state of phonons and a tritter.

\subsection{Two-mode squeezing}

To create a two-mode squeezed state of phonons, we require a Hamiltonian of the form (see e.g.\ \cite{CMF}):
\begin{align}
\hat{H} = r [ e^{i \vartheta_{sq} } \hat{b}\da_m \hat{b}\da_n + e^{-i \vartheta_{sq} } \hat{b}_m \hat{b}_n] .
\end{align}
This can be obtained from \eqref{eq:H} by choosing an oscillating potential to pick out these particular terms on resonance \cite{MassBEC}. For example $V_{\epsilon}(t) = \epsilon V_{0}  \cos \Omega t $ would achieve this where $\Omega := \omega_m + \omega_n$ and $V_{0}$ is a constant amplitude.  

\subsection{Tritter}

The Hamiltonian for a tritter is given by \eqref{eq:Htritter} where here we treat $\hat{a}_0$ as the annihilation operator for the condensate, and $\hat{a}_{n \neq 0}$ as the annihilation operator for the phonon modes, which we denoted as $\hat{b}_{n \neq 0}$ above. Since the condensate must be more populated than the phonon modes before and after the tritter for our description of the BEC used above to still hold, we can apply the Bogoliubov approximation and drop the hat on $\hat{a}_0$, leaving us with:
\begin{align}
H_{tr} = \frac{\hbar G}{\sqrt{2}} |a_0| \Big[ e^{i \vartheta} (\hat{b}_m + \hat{b}_n) + e^{-i \vartheta} \hat{a}_0 (\hat{b}^{\dagger}_m + \hat{b}^{\dagger}_n)  \Big].
\end{align}
This can be picked out from \eqref{eq:H} by choosing an oscillating potential of the form $V(t) = \epsilon V_0 \cos (\Omega t) \cos(\Omega' t)$, where $\Omega := \omega_m + \omega_n$ and $\Omega' := \omega_n - \omega_n$, and assuming that $\vartheta_n(\bs{r}) \approx \vartheta_m(\bs{r})$, which could be achieved, for example, by choosing modes with equal and opposite momenta in a uniform BEC with periodic boundary conditions \cite{PitaevskiiBook}.

\section{Heterodyne detection} \label{app:Heterodyne}

Rather than using a number-sum measurement, another possibility would be to use a  heterodyne measurement, for example, between the pump and the side modes.  Balanced homodyne detection for the side modes was considered in \cite{Li_2014} for a standard SU(1,1) interferometer and \cite{TruncatedSU11} for a `truncated' SU(1,1) experiment. In our considered heterodyne  case, at the measurement stage a balanced beam splitter could be applied between one of the side modes and the pump, and the difference of the number of particles in the two output parts of the final beam splitter could be considered: $\hat{S} = \hat{N}_1 - \hat{N}_2$. In the covariance matrix formalism we have:
\begin{align}
\braket{\hat{S}} &= \frac{1}{4} [ Tr(\sigma J_z) + d^T J_z d]\\
\mathrm{Var}(\hat{S}) &= \frac{1}{8} [Tr([\sigma J_z]^2 ) + 2 d^T J_z \sigma J_z d - 2n].
\end{align}
where:
\begin{align}
J_z = \ba{cc} \bs{1} & \bs{0} \\ \bs{0} & - \bs{1} \ea.
\end{align}
However, in order to measure the squeezing  parameter of the estimation channel, $\mathrm{Var}(\hat{S})$ or $\braket{\hat{S}^2}$ would need to be considered as the signal (see e.g.\ \cite{KnightBook}) and, therefore, the variance of this would be used in the error estimation.

\section{The quantum channels associated with a phononic BEC GW detector} \label{app:GWDetector}

In \cite{GWDetectorFirst}, the effect of a GW on phonons of a BEC is described using quantum field theory in curved spacetime. The GW is found to act as a unitary Bogoliubov transformation on the phononic field:\footnote{The BEC Bogoliubov transformations \eqref{eq:BogoTransP} were not considered in \cite{GWDetectorFirst} but are treated in \cite{MassBEC}.}
\begin{align} \label{eq:BogoTrans}
\hat{b}_m &:= \sum_n (A_{mn}^{\ast} \hat{b}_n + B^{\ast}_{mn} \hat{b}_n\da),
\end{align}
where $A_{mn}$ and $B_{mn}$ are Bogoliubov coefficients that depend on  the GW amplitude and frequency, and must also obey the following identities \cite{BirrelandDavies}:
\begin{align} \label{eq:cond1}
\sum_k [A_{ik} A^{\ast}_{jk}  - B_{ik} B^{\ast}_{jk} ] &= \delta_{ij},\\ \label{eq:cond2}
\sum_k [A_{ik} B_{jk} - B_{ik} A_{jk}] &= 0.
\end{align}	
This is a unitary transformation and, assuming Gaussian phonon states, the corresponding symplectic matrix  in the real $q,p$ basis is \cite{Accelerometer}:
\begin{align}
\bs{S} = \ba{ccccc} 
\bs{M_{11}} & \bs{M_{12}} & \hdots  & \hdots  & \hdots\\ 
\bs{M_{21}} & \ddots  & \ddots & \ddots & \ddots \\
\vdots  &  \ddots & \bs{M_{mm}} & \bs{M_{mn}}  & \ddots  \\
\vdots   & \ddots &  \bs{M_{nm}}  & \bs{M_{nn}} & \ddots   \\
\vdots   & \ddots &  \ddots  & \ddots & \ddots \ea ,
\end{align}
where:
\begin{align}
\bs{M_{mn}} = \ba{cc} \mathrm{Re}(A_{mn} - B_{mn}) & \mathrm{Im}(A_{mn} + B_{mn}) \\ - \mathrm{Im}(A_{mn} - B_{mn}) & \mathrm{Re}(A_{mn} + B_{mn}) \ea.
\end{align}
When the GW frequency $\Omega$ matches the sum of two phononic modes, $\Omega = \omega_m + \omega_n$, there is a resonant process such that, for example,  $B_{jk}$ was found to be \cite{GWDetectorFirst}:\footnote{Here we assume that $n \neq m$ such that we are considering two-mode squeezing. For a single-mode squeezing channel, see \cite{MassBEC}.} 
\begin{align} \label{eq:betajk}
B_{jk} (t)  = \frac{1}{4} \epsilon B e^{i \phi_{B}}(1 + (-1)^{j+k})\delta_{j+k,m+n}  + \mathcal{O} (\epsilon^2),
\end{align} 
where $\epsilon$ is the GW amplitude, $\phi_{B}$ is an arbitrary phase that we have included, and:
\begin{align} \label{eq:BGW}
B:=  \sqrt{\omega_j \omega_k} ct,
\end{align}
with $t$ the time of interaction between the GW and the phonons, $c$  a unitless constant due to the BEC Bogoliubov transformations \eqref{eq:BogoTrans} with the appropriate boundary conditions, and we have assumed a uniform BEC in a one-dimensional trap with the origin at the centre. There could also be resonances for the $A$-coefficients in this case \cite{JanThesis} but these would vanish when $j=m,k=n$ or $j=n,k=m$, which are the only cases of relevance here and discussed below.  When $\Omega = \omega_n - \omega_m$ (with $n > m$) there is also a resonant process such that, for example,  $A_{jk}$ is found to be:
\begin{align} \label{eq:alphajk}
A_{jk} (t)  = \frac{1}{4} A e^{i \phi_{A}} (1+ (-1)^{j+k}) (\delta_{j-k,n-m} - \delta_{j-k,m-n} ) + \mathcal{O} (\epsilon^2),
\end{align}
where $\phi_{A}$ is an arbitrary phase that we have included and:
\begin{align} \label{eq:AGW}
A:= \epsilon  \sqrt{\omega_j \omega_k} ct,
\end{align}
with $c$ is a unitless constant due to the BEC Bogoliubov transformations \eqref{eq:BogoTrans} with the appropriate boundary conditions. There may also be additional resonances for the $B$-coefficients in this case \cite{JanThesis} but these would vanish when $j=m,k=n$ or $j=n,k=m$, which are the only cases of relevance here and discussed below.

Considering a general phononic state with displacement and covariance matrices:
\begin{align}
\bs{d} &= (\bs{d}_1, \bs{d}_2, \ldots )^T,\\ 
\bs{\sigma_{mn}} &= \ba{cc} \bs{\psi_{mn}} & \bs{\phi_{mn}} \\ \bs{\phi^T_{mn}} & \bs{\psi'_{mn}} \ea,
\end{align} 
 then, after the GW acts on the phononic field, the state is given by:
\begin{align}
\bs{d} &= (\bs{d}'_1, \bs{d}'_2, \ldots )^T,\\ 
\bs{\sigma'_{mn}} &= \ba{cc} \bs{C_{mm}} & \bs{C_{mn}} \\ \bs{C_{nm}} & \bs{C_{nn}} \ea,
\end{align}
where:
\begin{align}
d_i &:= M_{ij} d_j,\\ 
C_{ij} &:= M^T_{ni} \psi_n M_{nj} + M^T_{mi} \phi^T_{nm} M_{nj} + M^T_{ni} \phi_{nm} M_{mj} + M^T_{mi} \psi_{m} M_{mj} + \sum_{l \neq m,n} M^T_{li} M_{lj}, 
\end{align}
and the last term is due to tracing out the other modes \cite{Accelerometer}. 

Since the phononic field is prepared such that modes $m$ and $n$ are in a large vacuum squeezed state, and all other modes are in the vacuum, when the GW has frequency $\Omega = \omega_m + \omega_n$, to second order in $\epsilon$, the action of the wave can then be effectively represented by the symplectic matrix:
\begin{align} \label{eq:SofGW}
\bs{S_{nm}} = \ba{cc} [1 + \frac{1}{2} s^2_{nm} ] \bs{1} & s_{nm} \bs{R_{\phi_{B}}} \\ s_{nm} \bs{R_{\phi_{B}}}&  [1 + \frac{1}{2} s^2_{nm} ] \bs{1} \ea,
\end{align}
where:
\begin{align}
\bs{R_{\phi_{B}}} &:= \ba{cc} \cos \phi_{B} & \sin \phi_{B } \\ \sin \phi_{B} & - \cos \phi_{B}\ea,\\  \label{eq:snm}
s_{nm}& = \frac{1}{4} \epsilon c \sqrt{\omega_m \omega_n}t ,
\end{align}
and we have used \eqref{eq:cond1}-\eqref{eq:cond2} up to second order in $\epsilon$. To second order also in the squeezing parameter $s$, \eqref{eq:SofGW} is simply a two-mode squeezing channel $U= e^{\xi \hat{b}_m\da \hat{b}_n\da - \xi^{\ast} \hat{b}_m \hat{b}_m}$ in modes $m$ and $n$:
\begin{align} \label{eq:SsGW}
\bs{S}_{s} &= \ba{cc}
\cosh s \bs{1} &  \sinh s \bs{R_{\phi_{B}}} \\
\sinh s \bs{R_{\phi_{B}}} &  \cosh s \bs{1} \ea,
\end{align}
where $\xi:= s e^{i \phi_{B}}$. This is as expected since the $B$-coefficients of a Bogoliubov transformation \eqref{eq:BogoTrans} are  associated with parametric down conversion. It is also clear from considering the GW as a time-varying gravitational field and applying the canonical approach presented in \cite{MassBEC}, which is essentially equivalent to choosing $\mathcal{V}_{\epsilon} (x,t) =  \epsilon m x^2 \Omega^2 \sin \Omega t / 4$ in \eqref{eq:Hint} for a quasi one-dimensional BEC analysed in the proper detector frame   \cite{GWSeismic}. From \cite{MassBEC,GWSeismic}, the unitless constant $c$ in \eqref{eq:betajk} is given by:
\begin{align} \label{eq:csq}
c:= \xi_n \xi_m \frac{n^2 + m^2}{(n-m)^2}, 
\end{align} 
where:
\begin{align}
\xi_n := \frac{m c^2_s}{\hbar \omega_n}
\end{align}
with $c_s$ the speed of sound of the BEC, and $\xi_n \gg 1$ for $\omega_n$ to correspond to a phononic frequency \cite{PitaevskiiBook}.

Analogous to the above, when the other resonance condition is chosen with $\Omega = \omega_n - \omega_m$, the action of the GW is effectively that of a mode-mixing channel  $U= e^{\zeta \hat{b}_m\da \hat{b}_n - \zeta^{\ast} \hat{b}_m \hat{b}_m\da}$ with $\zeta := s_{mn} e^{i \phi_{A}}$ up  to second order in $\epsilon$, and with $s_{mn}$ also given by \eqref{eq:snm}. In this case the symplectic matrix is given by:
\begin{align} \label{eq:SmGW}
\bs{S_{nm}} &= \ba{cc}
\cos  s_{nm} \bs{1} &  \sin s_{nm} \bs{R_{ \phi_{A}}} \\
-\sin s_{nm} \bs{R^T_{ \phi_{A}}} & \cos  s_{nm} \bs{1} \ea.
\end{align}
The fact that this is a mode-mixing channel is expected since the $A$-coefficients of a Bogoliubov transformation \eqref{eq:BogoTrans} are  associated with beam-splitting. It is also clear from considering the GW as a time-varying gravitational field and applying the canonical approach presented in \cite{MassBEC}. From \cite{MassBEC,GWSeismic}, the unitless constant $c$ in \eqref{eq:alphajk} is given by:
\begin{align} \label{eq:cm}
c:= \xi_n \xi_m \frac{n^2 + m^2}{(n+m)^2}.
\end{align}

\section*{Acknowledgements}

We thank Jan Kohlrus, Daniel Goldwater, Paul Juschitz, Tupac Bravo, Daniel Hartley and Dennis R\"{a}tzel for useful discussions and comments. R.H. and I.F. would like to acknowledge that this project was made possible through the support of the grant `Leaps in cosmology: gravitational wave detection with quantum systems' (No. 58745) from the John Templeton Foundation. The opinions expressed in this publication are those of the authors and do not necessarily reflect the views of the John Templeton Foundation.



\begin{thebibliography}{46}%
	\makeatletter
	\providecommand \@ifxundefined [1]{%
		\@ifx{#1\undefined}
	}%
	\providecommand \@ifnum [1]{%
		\ifnum #1\expandafter \@firstoftwo
		\else \expandafter \@secondoftwo
		\fi
	}%
	\providecommand \@ifx [1]{%
		\ifx #1\expandafter \@firstoftwo
		\else \expandafter \@secondoftwo
		\fi
	}%
	\providecommand \natexlab [1]{#1}%
	\providecommand \enquote  [1]{``#1''}%
	\providecommand \bibnamefont  [1]{#1}%
	\providecommand \bibfnamefont [1]{#1}%
	\providecommand \citenamefont [1]{#1}%
	\providecommand \href@noop [0]{\@secondoftwo}%
	\providecommand \href [0]{\begingroup \@sanitize@url \@href}%
	\providecommand \@href[1]{\@@startlink{#1}\@@href}%
	\providecommand \@@href[1]{\endgroup#1\@@endlink}%
	\providecommand \@sanitize@url [0]{\catcode `\\12\catcode `\$12\catcode
		`\&12\catcode `\#12\catcode `\^12\catcode `\_12\catcode `\%12\relax}%
	\providecommand \@@startlink[1]{}%
	\providecommand \@@endlink[0]{}%
	\providecommand \url  [0]{\begingroup\@sanitize@url \@url }%
	\providecommand \@url [1]{\endgroup\@href {#1}{\urlprefix }}%
	\providecommand \urlprefix  [0]{URL }%
	\providecommand \Eprint [0]{\href }%
	\providecommand \doibase [0]{http://dx.doi.org/}%
	\providecommand \selectlanguage [0]{\@gobble}%
	\providecommand \bibinfo  [0]{\@secondoftwo}%
	\providecommand \bibfield  [0]{\@secondoftwo}%
	\providecommand \translation [1]{[#1]}%
	\providecommand \BibitemOpen [0]{}%
	\providecommand \bibitemStop [0]{}%
	\providecommand \bibitemNoStop [0]{.\EOS\space}%
	\providecommand \EOS [0]{\spacefactor3000\relax}%
	\providecommand \BibitemShut  [1]{\csname bibitem#1\endcsname}%
	\let\auto@bib@innerbib\@empty
	\bibitem [{\citenamefont {Abbott}\ \emph {et~al.}(2016)\citenamefont {Abbott}
		\emph {et~al.}}]{GWDetection}%
	\BibitemOpen
	\bibfield  {author} {\bibinfo {author} {\bibfnamefont {B.~P.}\ \bibnamefont
			{Abbott}} \emph {et~al.} (\bibinfo {collaboration} {LIGO Scientific
			Collaboration and Virgo Collaboration}),\ }\href {\doibase
		10.1103/PhysRevLett.116.061102} {\bibfield  {journal} {\bibinfo  {journal}
			{Phys. Rev. Lett.}\ }\textbf {\bibinfo {volume} {116}},\ \bibinfo {pages}
		{061102} (\bibinfo {year} {2016})}\BibitemShut {NoStop}%
	\bibitem [{\citenamefont {Aasi}\ \emph {et~al.}(2013)\citenamefont {Aasi} \emph
		{et~al.}}]{aasi2013enhanced}%
	\BibitemOpen
	\bibfield  {author} {\bibinfo {author} {\bibfnamefont {J.}~\bibnamefont
			{Aasi}} \emph {et~al.},\ }\href {\doibase 10.1038/nphoton.2013.177}
	{\bibfield  {journal} {\bibinfo  {journal} {Nature Photonics}\ }\textbf
		{\bibinfo {volume} {7}},\ \bibinfo {pages} {613} (\bibinfo {year}
		{2013})}\BibitemShut {NoStop}%
	\bibitem [{\citenamefont {Olivares}\ and\ \citenamefont
		{Paris}(2007)}]{Olivares2007}%
	\BibitemOpen
	\bibfield  {author} {\bibinfo {author} {\bibfnamefont {S.}~\bibnamefont
			{Olivares}}\ and\ \bibinfo {author} {\bibfnamefont {M.~G.~A.}\ \bibnamefont
			{Paris}},\ }\href {\doibase 10.1134/S0030400X07080103} {\bibfield  {journal}
		{\bibinfo  {journal} {Optics and Spectroscopy}\ }\textbf {\bibinfo {volume}
			{103}},\ \bibinfo {pages} {231} (\bibinfo {year} {2007})}\BibitemShut
	{NoStop}%
	\bibitem [{\citenamefont {Yurke}\ \emph {et~al.}(1986)\citenamefont {Yurke},
		\citenamefont {McCall},\ and\ \citenamefont {Klauder}}]{SU11}%
	\BibitemOpen
	\bibfield  {author} {\bibinfo {author} {\bibfnamefont {B.}~\bibnamefont
			{Yurke}}, \bibinfo {author} {\bibfnamefont {S.~L.}\ \bibnamefont {McCall}}, \
		and\ \bibinfo {author} {\bibfnamefont {J.~R.}\ \bibnamefont {Klauder}},\
	}\href {\doibase 10.1103/PhysRevA.33.4033} {\bibfield  {journal} {\bibinfo
			{journal} {Phys. Rev. A}\ }\textbf {\bibinfo {volume} {33}},\ \bibinfo
		{pages} {4033} (\bibinfo {year} {1986})}\BibitemShut {NoStop}%
	\bibitem [{\citenamefont {Jing}\ \emph {et~al.}(2011)\citenamefont {Jing},
		\citenamefont {Liu}, \citenamefont {Zhou}, \citenamefont {Ou},\ and\
		\citenamefont {Zhang}}]{SU11Optical1}%
	\BibitemOpen
	\bibfield  {author} {\bibinfo {author} {\bibfnamefont {J.}~\bibnamefont
			{Jing}}, \bibinfo {author} {\bibfnamefont {C.}~\bibnamefont {Liu}}, \bibinfo
		{author} {\bibfnamefont {Z.}~\bibnamefont {Zhou}}, \bibinfo {author}
		{\bibfnamefont {Z.~Y.}\ \bibnamefont {Ou}}, \ and\ \bibinfo {author}
		{\bibfnamefont {W.}~\bibnamefont {Zhang}},\ }\href {\doibase
		10.1063/1.3606549} {\bibfield  {journal} {\bibinfo  {journal} {Applied
				Physics Letters}\ }\textbf {\bibinfo {volume} {99}},\ \bibinfo {pages}
		{011110} (\bibinfo {year} {2011})}\BibitemShut {NoStop}%
	\bibitem [{\citenamefont {Hudelist}\ \emph {et~al.}(2014)\citenamefont
		{Hudelist}, \citenamefont {Kong}, \citenamefont {Liu}, \citenamefont {Jing},
		\citenamefont {Ou},\ and\ \citenamefont {Zhang}}]{SU11Optical2}%
	\BibitemOpen
	\bibfield  {author} {\bibinfo {author} {\bibfnamefont {F.}~\bibnamefont
			{Hudelist}}, \bibinfo {author} {\bibfnamefont {J.}~\bibnamefont {Kong}},
		\bibinfo {author} {\bibfnamefont {C.}~\bibnamefont {Liu}}, \bibinfo {author}
		{\bibfnamefont {J.}~\bibnamefont {Jing}}, \bibinfo {author} {\bibfnamefont
			{Z.}~\bibnamefont {Ou}}, \ and\ \bibinfo {author} {\bibfnamefont
			{W.}~\bibnamefont {Zhang}},\ }\href {\doibase 10.1038/ncomms4049} {\bibfield
		{journal} {\bibinfo  {journal} {Nature communications}\ }\textbf {\bibinfo
			{volume} {5}},\ \bibinfo {pages} {3049} (\bibinfo {year} {2014})}\BibitemShut
	{NoStop}%
	\bibitem [{\citenamefont {Chen}\ \emph {et~al.}(2015)\citenamefont {Chen},
		\citenamefont {Qiu}, \citenamefont {Chen}, \citenamefont {Guo}, \citenamefont
		{Chen}, \citenamefont {Ou},\ and\ \citenamefont {Zhang}}]{SU11AtomLight}%
	\BibitemOpen
	\bibfield  {author} {\bibinfo {author} {\bibfnamefont {B.}~\bibnamefont
			{Chen}}, \bibinfo {author} {\bibfnamefont {C.}~\bibnamefont {Qiu}}, \bibinfo
		{author} {\bibfnamefont {S.}~\bibnamefont {Chen}}, \bibinfo {author}
		{\bibfnamefont {J.}~\bibnamefont {Guo}}, \bibinfo {author} {\bibfnamefont
			{L.~Q.}\ \bibnamefont {Chen}}, \bibinfo {author} {\bibfnamefont {Z.~Y.}\
			\bibnamefont {Ou}}, \ and\ \bibinfo {author} {\bibfnamefont {W.}~\bibnamefont
			{Zhang}},\ }\href {\doibase 10.1103/PhysRevLett.115.043602} {\bibfield
		{journal} {\bibinfo  {journal} {Phys. Rev. Lett.}\ }\textbf {\bibinfo
			{volume} {115}},\ \bibinfo {pages} {043602} (\bibinfo {year}
		{2015})}\BibitemShut {NoStop}%
	\bibitem [{\citenamefont {Gross}\ \emph {et~al.}(2012)\citenamefont {Gross},
		\citenamefont {Zibold}, \citenamefont {Nicklas}, \citenamefont {Esteve},\
		and\ \citenamefont {Oberthaler}}]{SU11BEC1}%
	\BibitemOpen
	\bibfield  {author} {\bibinfo {author} {\bibfnamefont {C.}~\bibnamefont
			{Gross}}, \bibinfo {author} {\bibfnamefont {T.}~\bibnamefont {Zibold}},
		\bibinfo {author} {\bibfnamefont {E.}~\bibnamefont {Nicklas}}, \bibinfo
		{author} {\bibfnamefont {J.}~\bibnamefont {Esteve}}, \ and\ \bibinfo {author}
		{\bibfnamefont {M.~K.}\ \bibnamefont {Oberthaler}},\ }\href {\doibase
		10.1038/nature08919} {\bibfield  {journal} {\bibinfo  {journal} {Nature}\
		}\textbf {\bibinfo {volume} {464}},\ \bibinfo {pages} {1165} (\bibinfo {year}
		{2012})}\BibitemShut {NoStop}%
	\bibitem [{\citenamefont {Linnemann}\ \emph {et~al.}(2016)\citenamefont
		{Linnemann}, \citenamefont {Strobel}, \citenamefont {Muessel}, \citenamefont
		{Schulz}, \citenamefont {Lewis-Swan}, \citenamefont {Kheruntsyan},\ and\
		\citenamefont {Oberthaler}}]{SU11BEC3}%
	\BibitemOpen
	\bibfield  {author} {\bibinfo {author} {\bibfnamefont {D.}~\bibnamefont
			{Linnemann}}, \bibinfo {author} {\bibfnamefont {H.}~\bibnamefont {Strobel}},
		\bibinfo {author} {\bibfnamefont {W.}~\bibnamefont {Muessel}}, \bibinfo
		{author} {\bibfnamefont {J.}~\bibnamefont {Schulz}}, \bibinfo {author}
		{\bibfnamefont {R.~J.}\ \bibnamefont {Lewis-Swan}}, \bibinfo {author}
		{\bibfnamefont {K.~V.}\ \bibnamefont {Kheruntsyan}}, \ and\ \bibinfo {author}
		{\bibfnamefont {M.~K.}\ \bibnamefont {Oberthaler}},\ }\href {\doibase
		10.1103/PhysRevLett.117.013001} {\bibfield  {journal} {\bibinfo  {journal}
			{Phys. Rev. Lett.}\ }\textbf {\bibinfo {volume} {117}},\ \bibinfo {pages}
		{013001} (\bibinfo {year} {2016})}\BibitemShut {NoStop}%
	\bibitem [{\citenamefont {Szigeti}\ \emph {et~al.}(2017)\citenamefont
		{Szigeti}, \citenamefont {Lewis-Swan},\ and\ \citenamefont
		{Haine}}]{PumpedUpSU11}%
	\BibitemOpen
	\bibfield  {author} {\bibinfo {author} {\bibfnamefont {S.~S.}\ \bibnamefont
			{Szigeti}}, \bibinfo {author} {\bibfnamefont {R.~J.}\ \bibnamefont
			{Lewis-Swan}}, \ and\ \bibinfo {author} {\bibfnamefont {S.~A.}\ \bibnamefont
			{Haine}},\ }\href {\doibase 10.1103/PhysRevLett.118.150401} {\bibfield
		{journal} {\bibinfo  {journal} {Phys. Rev. Lett.}\ }\textbf {\bibinfo
			{volume} {118}},\ \bibinfo {pages} {150401} (\bibinfo {year}
		{2017})}\BibitemShut {NoStop}%
	\bibitem [{\citenamefont {Demkowicz-Dobrzański}\ \emph
		{et~al.}(2015)\citenamefont {Demkowicz-Dobrzański}, \citenamefont
		{Jarzyna},\ and\ \citenamefont {Kołodyński}}]{QuantumLimitsOptics}%
	\BibitemOpen
	\bibfield  {author} {\bibinfo {author} {\bibfnamefont {R.}~\bibnamefont
			{Demkowicz-Dobrzański}}, \bibinfo {author} {\bibfnamefont {M.}~\bibnamefont
			{Jarzyna}}, \ and\ \bibinfo {author} {\bibfnamefont {J.}~\bibnamefont
			{Kołodyński}}\ }(\bibinfo  {publisher} {Elsevier},\ \bibinfo {year}
	{2015})\ pp.\ \bibinfo {pages} {345 -- 435}\BibitemShut {NoStop}%
	\bibitem [{\citenamefont {Helstrom}(1969)}]{CramerRao1}%
	\BibitemOpen
	\bibfield  {author} {\bibinfo {author} {\bibfnamefont {C.~W.}\ \bibnamefont
			{Helstrom}},\ }\href {\doibase 10.1007/BF01007479} {\bibfield  {journal}
		{\bibinfo  {journal} {Journal of Statistical Physics}\ }\textbf {\bibinfo
			{volume} {1}},\ \bibinfo {pages} {231} (\bibinfo {year} {1969})}\BibitemShut
	{NoStop}%
	\bibitem [{\citenamefont {Kay}(1993)}]{CramerRao2}%
	\BibitemOpen
	\bibfield  {author} {\bibinfo {author} {\bibfnamefont {S.~M.}\ \bibnamefont
			{Kay}},\ }\href@noop {} {\emph {\bibinfo {title} {Fundamentals of Statistical
				Signal Processing, Volume I: Estimation Theory}}},\ edited by\ \bibinfo
	{editor} {\bibfnamefont {A.~V.}\ \bibnamefont {Oppenheim}}\ (\bibinfo {year}
	{1993})\BibitemShut {NoStop}%
	\bibitem [{\citenamefont {Braunstein}\ and\ \citenamefont {Caves}(1994)}]{QFI}%
	\BibitemOpen
	\bibfield  {author} {\bibinfo {author} {\bibfnamefont {S.~L.}\ \bibnamefont
			{Braunstein}}\ and\ \bibinfo {author} {\bibfnamefont {C.~M.}\ \bibnamefont
			{Caves}},\ }\href {\doibase 10.1103/PhysRevLett.72.3439} {\bibfield
		{journal} {\bibinfo  {journal} {Phys. Rev. Lett.}\ }\textbf {\bibinfo
			{volume} {72}},\ \bibinfo {pages} {3439} (\bibinfo {year}
		{1994})}\BibitemShut {NoStop}%
	\bibitem [{\citenamefont {{Sab{\'{\i}}n}}\ \emph {et~al.}(2014)\citenamefont
		{{Sab{\'{\i}}n}}, \citenamefont {{Bruschi}}, \citenamefont {{Ahmadi}},\ and\
		\citenamefont {{Fuentes}}}]{GWDetectorFirst}%
	\BibitemOpen
	\bibfield  {author} {\bibinfo {author} {\bibfnamefont {C.}~\bibnamefont
			{{Sab{\'{\i}}n}}}, \bibinfo {author} {\bibfnamefont {D.~E.}\ \bibnamefont
			{{Bruschi}}}, \bibinfo {author} {\bibfnamefont {M.}~\bibnamefont {{Ahmadi}}},
		\ and\ \bibinfo {author} {\bibfnamefont {I.}~\bibnamefont {{Fuentes}}},\
	}\href {\doibase 10.1088/1367-2630/16/8/085003} {\bibfield  {journal}
		{\bibinfo  {journal} {New Journal of Physics}\ }\textbf {\bibinfo {volume}
			{16}},\ \bibinfo {eid} {085003} (\bibinfo {year} {2014})}\BibitemShut
	{NoStop}%
	\bibitem [{\citenamefont {{Ferraro}}\ \emph {et~al.}(2005)\citenamefont
		{{Ferraro}}, \citenamefont {{Olivares}},\ and\ \citenamefont
		{{Paris}}}]{CMF}%
	\BibitemOpen
	\bibfield  {author} {\bibinfo {author} {\bibfnamefont {A.}~\bibnamefont
			{{Ferraro}}}, \bibinfo {author} {\bibfnamefont {S.}~\bibnamefont
			{{Olivares}}}, \ and\ \bibinfo {author} {\bibfnamefont {M.~G.~A.}\
			\bibnamefont {{Paris}}},\ }\href@noop {} {\  (\bibinfo {year} {2005})},\
	\Eprint {http://arxiv.org/abs/quant-ph/0503237} {arXiv:quant-ph/0503237}
	\BibitemShut {NoStop}%
	\bibitem [{\citenamefont {\ifmmode~\check{S}\else \v{S}\fi{}afr\'anek}\ and\
		\citenamefont {Fuentes}(2016)}]{DomOptimum}%
	\BibitemOpen
	\bibfield  {author} {\bibinfo {author} {\bibfnamefont {D.}~\bibnamefont
			{\ifmmode~\check{S}\else \v{S}\fi{}afr\'anek}}\ and\ \bibinfo {author}
		{\bibfnamefont {I.}~\bibnamefont {Fuentes}},\ }\href {\doibase
		10.1103/PhysRevA.94.062313} {\bibfield  {journal} {\bibinfo  {journal} {Phys.
				Rev. A}\ }\textbf {\bibinfo {volume} {94}},\ \bibinfo {pages} {062313}
		(\bibinfo {year} {2016})}\BibitemShut {NoStop}%
	\bibitem [{\citenamefont {{Monras}}(2013)}]{QFIMatrices}%
	\BibitemOpen
	\bibfield  {author} {\bibinfo {author} {\bibfnamefont {A.}~\bibnamefont
			{{Monras}}},\ }\href@noop {} {\bibfield  {journal} {\bibinfo  {journal}
			{ArXiv e-prints}\ } (\bibinfo {year} {2013})},\ \Eprint
	{http://arxiv.org/abs/1303.3682} {arXiv:1303.3682 [quant-ph]} \BibitemShut
	{NoStop}%
	\bibitem [{\citenamefont {Pinel}\ \emph {et~al.}(2013)\citenamefont {Pinel},
		\citenamefont {Jian}, \citenamefont {Treps}, \citenamefont {Fabre},\ and\
		\citenamefont {Braun}}]{QFIMatricesBraun}%
	\BibitemOpen
	\bibfield  {author} {\bibinfo {author} {\bibfnamefont {O.}~\bibnamefont
			{Pinel}}, \bibinfo {author} {\bibfnamefont {P.}~\bibnamefont {Jian}},
		\bibinfo {author} {\bibfnamefont {N.}~\bibnamefont {Treps}}, \bibinfo
		{author} {\bibfnamefont {C.}~\bibnamefont {Fabre}}, \ and\ \bibinfo {author}
		{\bibfnamefont {D.}~\bibnamefont {Braun}},\ }\href {\doibase
		10.1103/PhysRevA.88.040102} {\bibfield  {journal} {\bibinfo  {journal} {Phys.
				Rev. A}\ }\textbf {\bibinfo {volume} {88}},\ \bibinfo {pages} {040102}
		(\bibinfo {year} {2013})}\BibitemShut {NoStop}%
	\bibitem [{\citenamefont {Arvind}\ \emph {et~al.}(1995)\citenamefont {Arvind},
		\citenamefont {Dutta}, \citenamefont {Mukunda},\ and\ \citenamefont
		{Simon}}]{Arvind1995}%
	\BibitemOpen
	\bibfield  {author} {\bibinfo {author} {\bibnamefont {Arvind}}, \bibinfo
		{author} {\bibfnamefont {B.}~\bibnamefont {Dutta}}, \bibinfo {author}
		{\bibfnamefont {N.}~\bibnamefont {Mukunda}}, \ and\ \bibinfo {author}
		{\bibfnamefont {R.}~\bibnamefont {Simon}},\ }\href {\doibase
		10.1007/BF02848172} {\bibfield  {journal} {\bibinfo  {journal} {Pramana}\
		}\textbf {\bibinfo {volume} {45}},\ \bibinfo {pages} {471} (\bibinfo {year}
		{1995})}\BibitemShut {NoStop}%
	\bibitem [{\citenamefont {Sparaciari}\ \emph {et~al.}(2016)\citenamefont
		{Sparaciari}, \citenamefont {Olivares},\ and\ \citenamefont
		{Paris}}]{GaussianActive}%
	\BibitemOpen
	\bibfield  {author} {\bibinfo {author} {\bibfnamefont {C.}~\bibnamefont
			{Sparaciari}}, \bibinfo {author} {\bibfnamefont {S.}~\bibnamefont
			{Olivares}}, \ and\ \bibinfo {author} {\bibfnamefont {M.~G.~A.}\ \bibnamefont
			{Paris}},\ }\href {\doibase 10.1103/PhysRevA.93.023810} {\bibfield  {journal}
		{\bibinfo  {journal} {Phys. Rev. A}\ }\textbf {\bibinfo {volume} {93}},\
		\bibinfo {pages} {023810} (\bibinfo {year} {2016})}\BibitemShut {NoStop}%
	\bibitem [{\citenamefont {Weber}(1966)}]{WeberBar}%
	\BibitemOpen
	\bibfield  {author} {\bibinfo {author} {\bibfnamefont {J.}~\bibnamefont
			{Weber}},\ }\href {\doibase 10.1103/PhysRevLett.17.1228} {\bibfield
		{journal} {\bibinfo  {journal} {Phys. Rev. Lett.}\ }\textbf {\bibinfo
			{volume} {17}},\ \bibinfo {pages} {1228} (\bibinfo {year}
		{1966})}\BibitemShut {NoStop}%
	\bibitem [{\citenamefont {Howl}\ \emph {et~al.}(2018)\citenamefont {Howl},
		\citenamefont {Hackerm\"{u}ller}, \citenamefont {Bruschi},\ and\
		\citenamefont {Fuentes}}]{ReviewPaper}%
	\BibitemOpen
	\bibfield  {author} {\bibinfo {author} {\bibfnamefont {R.}~\bibnamefont
			{Howl}}, \bibinfo {author} {\bibfnamefont {L.}~\bibnamefont
			{Hackerm\"{u}ller}}, \bibinfo {author} {\bibfnamefont {D.~E.}\ \bibnamefont
			{Bruschi}}, \ and\ \bibinfo {author} {\bibfnamefont {I.}~\bibnamefont
			{Fuentes}},\ }\href {\doibase 10.1080/23746149.2017.1383184} {\bibfield
		{journal} {\bibinfo  {journal} {Advances in Physics: X}\ }\textbf {\bibinfo
			{volume} {3}},\ \bibinfo {pages} {1383184} (\bibinfo {year}
		{2018})}\BibitemShut {NoStop}%
	\bibitem [{\citenamefont {Braginskii}\ \emph {et~al.}(1973)\citenamefont
		{Braginskii}, \citenamefont {Grishchuk}, \citenamefont {Doroshkevich},
		\citenamefont {Zeldovich}, \citenamefont {Novikov},\ and\ \citenamefont
		{Sazhin}}]{braginskii1974electromagnetic}%
	\BibitemOpen
	\bibfield  {author} {\bibinfo {author} {\bibfnamefont {V.}~\bibnamefont
			{Braginskii}}, \bibinfo {author} {\bibfnamefont {L.}~\bibnamefont
			{Grishchuk}}, \bibinfo {author} {\bibfnamefont {A.}~\bibnamefont
			{Doroshkevich}}, \bibinfo {author} {\bibfnamefont {I.~B.}\ \bibnamefont
			{Zeldovich}}, \bibinfo {author} {\bibfnamefont {I.}~\bibnamefont {Novikov}},
		\ and\ \bibinfo {author} {\bibfnamefont {M.}~\bibnamefont {Sazhin}},\
	}\href@noop {} {\bibfield  {journal} {\bibinfo  {journal} {Zhurnal
				Eksperimentalnoi i Teoreticheskoi Fiziki}\ }\textbf {\bibinfo {volume}
			{65}},\ \bibinfo {pages} {1729} (\bibinfo {year} {1973})}\BibitemShut
	{NoStop}%
	\bibitem [{\citenamefont {Rovelli}\ and\ \citenamefont
		{Vidotto}(2014)}]{PlanckStars}%
	\BibitemOpen
	\bibfield  {author} {\bibinfo {author} {\bibfnamefont {C.}~\bibnamefont
			{Rovelli}}\ and\ \bibinfo {author} {\bibfnamefont {F.}~\bibnamefont
			{Vidotto}},\ }\href {\doibase 10.1142/S0218271814420267} {\bibfield
		{journal} {\bibinfo  {journal} {International Journal of Modern Physics D}\
		}\textbf {\bibinfo {volume} {23}},\ \bibinfo {pages} {1442026} (\bibinfo
		{year} {2014})}\BibitemShut {NoStop}%
	\bibitem [{\citenamefont {Carusotto}\ \emph {et~al.}(2010)\citenamefont
		{Carusotto}, \citenamefont {Balbinot}, \citenamefont {Fabbri},\ and\
		\citenamefont {Recati}}]{DCETheory}%
	\BibitemOpen
	\bibfield  {author} {\bibinfo {author} {\bibfnamefont {I.}~\bibnamefont
			{Carusotto}}, \bibinfo {author} {\bibfnamefont {R.}~\bibnamefont {Balbinot}},
		\bibinfo {author} {\bibfnamefont {A.}~\bibnamefont {Fabbri}}, \ and\ \bibinfo
		{author} {\bibfnamefont {A.}~\bibnamefont {Recati}},\ }\href {\doibase
		10.1140/epjd/e2009-00314-3} {\bibfield  {journal} {\bibinfo  {journal} {The
				European Physical Journal D}\ }\textbf {\bibinfo {volume} {56}},\ \bibinfo
		{pages} {391} (\bibinfo {year} {2010})}\BibitemShut {NoStop}%
	\bibitem [{\citenamefont {Jaskula}\ \emph {et~al.}(2012)\citenamefont
		{Jaskula}, \citenamefont {Partridge}, \citenamefont {Bonneau}, \citenamefont
		{Lopes}, \citenamefont {Ruaudel}, \citenamefont {Boiron},\ and\ \citenamefont
		{Westbrook}}]{DCEWestbrook}%
	\BibitemOpen
	\bibfield  {author} {\bibinfo {author} {\bibfnamefont {J.-C.}\ \bibnamefont
			{Jaskula}}, \bibinfo {author} {\bibfnamefont {G.~B.}\ \bibnamefont
			{Partridge}}, \bibinfo {author} {\bibfnamefont {M.}~\bibnamefont {Bonneau}},
		\bibinfo {author} {\bibfnamefont {R.}~\bibnamefont {Lopes}}, \bibinfo
		{author} {\bibfnamefont {J.}~\bibnamefont {Ruaudel}}, \bibinfo {author}
		{\bibfnamefont {D.}~\bibnamefont {Boiron}}, \ and\ \bibinfo {author}
		{\bibfnamefont {C.~I.}\ \bibnamefont {Westbrook}},\ }\href {\doibase
		10.1103/PhysRevLett.109.220401} {\bibfield  {journal} {\bibinfo  {journal}
			{Phys. Rev. Lett.}\ }\textbf {\bibinfo {volume} {109}},\ \bibinfo {pages}
		{220401} (\bibinfo {year} {2012})}\BibitemShut {NoStop}%
	\bibitem [{\citenamefont {{Ahmadi}}\ \emph {et~al.}(2014)\citenamefont
		{{Ahmadi}}, \citenamefont {{Bruschi}}, \citenamefont {{Sab{\'{\i}}n}},
		\citenamefont {{Adesso}},\ and\ \citenamefont {{Fuentes}}}]{Accelerometer}%
	\BibitemOpen
	\bibfield  {author} {\bibinfo {author} {\bibfnamefont {M.}~\bibnamefont
			{{Ahmadi}}}, \bibinfo {author} {\bibfnamefont {D.~E.}\ \bibnamefont
			{{Bruschi}}}, \bibinfo {author} {\bibfnamefont {C.}~\bibnamefont
			{{Sab{\'{\i}}n}}}, \bibinfo {author} {\bibfnamefont {G.}~\bibnamefont
			{{Adesso}}}, \ and\ \bibinfo {author} {\bibfnamefont {I.}~\bibnamefont
			{{Fuentes}}},\ }\href {\doibase 10.1038/srep04996} {\bibfield  {journal}
		{\bibinfo  {journal} {Scientific Reports}\ }\textbf {\bibinfo {volume} {4}},\
		\bibinfo {eid} {4996} (\bibinfo {year} {2014})}\BibitemShut {NoStop}%
	\bibitem [{\citenamefont {{Michael}}\ \emph {et~al.}(2018)\citenamefont
		{{Michael}}, \citenamefont {{Schmiedmayer}},\ and\ \citenamefont
		{{Demler}}}]{SchmiedmayerDCE}%
	\BibitemOpen
	\bibfield  {author} {\bibinfo {author} {\bibfnamefont {M.~H.}\ \bibnamefont
			{{Michael}}}, \bibinfo {author} {\bibfnamefont {J.}~\bibnamefont
			{{Schmiedmayer}}}, \ and\ \bibinfo {author} {\bibfnamefont {E.}~\bibnamefont
			{{Demler}}},\ }\href@noop {} {\  (\bibinfo {year} {2018})},\ \Eprint
	{http://arxiv.org/abs/1812.05114} {arXiv:1812.05114} \BibitemShut {NoStop}%
	\bibitem [{\citenamefont {R\"{a}tzel}\ \emph {et~al.}(2018)\citenamefont
		{R\"{a}tzel}, \citenamefont {Howl}, \citenamefont {Lindkvist},\ and\
		\citenamefont {Fuentes}}]{MassBEC}%
	\BibitemOpen
	\bibfield  {author} {\bibinfo {author} {\bibfnamefont {D.}~\bibnamefont
			{R\"{a}tzel}}, \bibinfo {author} {\bibfnamefont {R.}~\bibnamefont {Howl}},
		\bibinfo {author} {\bibfnamefont {J.}~\bibnamefont {Lindkvist}}, \ and\
		\bibinfo {author} {\bibfnamefont {I.}~\bibnamefont {Fuentes}},\ }\href
	{\doibase 10.1088/1367-2630/aad272} {\bibfield  {journal} {\bibinfo
			{journal} {New Journal of Physics}\ }\textbf {\bibinfo {volume} {20}},\
		\bibinfo {pages} {073044} (\bibinfo {year} {2018})}\BibitemShut {NoStop}%
	\bibitem [{\citenamefont {Wade}\ \emph {et~al.}(2015)\citenamefont {Wade},
		\citenamefont {Sherson},\ and\ \citenamefont
		{M\o{}lmer}}]{PhysRevLett.115.060401}%
	\BibitemOpen
	\bibfield  {author} {\bibinfo {author} {\bibfnamefont {A.~C.~J.}\
			\bibnamefont {Wade}}, \bibinfo {author} {\bibfnamefont {J.~F.}\ \bibnamefont
			{Sherson}}, \ and\ \bibinfo {author} {\bibfnamefont {K.}~\bibnamefont
			{M\o{}lmer}},\ }\href {\doibase 10.1103/PhysRevLett.115.060401} {\bibfield
		{journal} {\bibinfo  {journal} {Phys. Rev. Lett.}\ }\textbf {\bibinfo
			{volume} {115}},\ \bibinfo {pages} {060401} (\bibinfo {year}
		{2015})}\BibitemShut {NoStop}%
	\bibitem [{\citenamefont {Wade}\ \emph {et~al.}(2016)\citenamefont {Wade},
		\citenamefont {Sherson},\ and\ \citenamefont
		{M\o{}lmer}}]{PhysRevA.93.023610}%
	\BibitemOpen
	\bibfield  {author} {\bibinfo {author} {\bibfnamefont {A.~C.~J.}\
			\bibnamefont {Wade}}, \bibinfo {author} {\bibfnamefont {J.~F.}\ \bibnamefont
			{Sherson}}, \ and\ \bibinfo {author} {\bibfnamefont {K.}~\bibnamefont
			{M\o{}lmer}},\ }\href {\doibase 10.1103/PhysRevA.93.023610} {\bibfield
		{journal} {\bibinfo  {journal} {Phys. Rev. A}\ }\textbf {\bibinfo {volume}
			{93}},\ \bibinfo {pages} {023610} (\bibinfo {year} {2016})}\BibitemShut
	{NoStop}%
	\bibitem [{\citenamefont {Steinhauer}(2016)}]{Steinhauer2016}%
	\BibitemOpen
	\bibfield  {author} {\bibinfo {author} {\bibfnamefont {J.}~\bibnamefont
			{Steinhauer}},\ }\href {http://dx.doi.org/10.1038/nphys3863} {\bibfield
		{journal} {\bibinfo  {journal} {Nat Phys}\ }\textbf {\bibinfo {volume}
			{12}},\ \bibinfo {pages} {959} (\bibinfo {year} {2016})}\BibitemShut
	{NoStop}%
	\bibitem [{\citenamefont {Rogel-Salazar}\ \emph {et~al.}(2002)\citenamefont
		{Rogel-Salazar}, \citenamefont {New}, \citenamefont {Choi},\ and\
		\citenamefont {Burnett}}]{BeliaevEntanglement}%
	\BibitemOpen
	\bibfield  {author} {\bibinfo {author} {\bibfnamefont {J.}~\bibnamefont
			{Rogel-Salazar}}, \bibinfo {author} {\bibfnamefont {G.~H.~C.}\ \bibnamefont
			{New}}, \bibinfo {author} {\bibfnamefont {S.}~\bibnamefont {Choi}}, \ and\
		\bibinfo {author} {\bibfnamefont {K.}~\bibnamefont {Burnett}},\ }\href
	{\doibase 10.1103/PhysRevA.65.023601} {\bibfield  {journal} {\bibinfo
			{journal} {Phys. Rev. A}\ }\textbf {\bibinfo {volume} {65}},\ \bibinfo
		{pages} {023601} (\bibinfo {year} {2002})}\BibitemShut {NoStop}%
	\bibitem [{\citenamefont {Tozzo}\ and\ \citenamefont
		{Dalfovo}(2004)}]{PhononEvap}%
	\BibitemOpen
	\bibfield  {author} {\bibinfo {author} {\bibfnamefont {C.}~\bibnamefont
			{Tozzo}}\ and\ \bibinfo {author} {\bibfnamefont {F.}~\bibnamefont
			{Dalfovo}},\ }\href {\doibase 10.1103/PhysRevA.69.053606} {\bibfield
		{journal} {\bibinfo  {journal} {Phys. Rev. A}\ }\textbf {\bibinfo {volume}
			{69}},\ \bibinfo {pages} {053606} (\bibinfo {year} {2004})}\BibitemShut
	{NoStop}%
	\bibitem [{\citenamefont {Katz}\ \emph {et~al.}(2004)\citenamefont {Katz},
		\citenamefont {Ozeri}, \citenamefont {Steinhauer}, \citenamefont {Davidson},
		\citenamefont {Tozzo},\ and\ \citenamefont {Dalfovo}}]{HeterodyneBECs}%
	\BibitemOpen
	\bibfield  {author} {\bibinfo {author} {\bibfnamefont {N.}~\bibnamefont
			{Katz}}, \bibinfo {author} {\bibfnamefont {R.}~\bibnamefont {Ozeri}},
		\bibinfo {author} {\bibfnamefont {J.}~\bibnamefont {Steinhauer}}, \bibinfo
		{author} {\bibfnamefont {N.}~\bibnamefont {Davidson}}, \bibinfo {author}
		{\bibfnamefont {C.}~\bibnamefont {Tozzo}}, \ and\ \bibinfo {author}
		{\bibfnamefont {F.}~\bibnamefont {Dalfovo}},\ }\href {\doibase
		10.1103/PhysRevLett.93.220403} {\bibfield  {journal} {\bibinfo  {journal}
			{Phys. Rev. Lett.}\ }\textbf {\bibinfo {volume} {93}},\ \bibinfo {pages}
		{220403} (\bibinfo {year} {2004})}\BibitemShut {NoStop}%
	\bibitem [{\citenamefont {Maggiore}(2008)}]{maggiore2008gravitational}%
	\BibitemOpen
	\bibfield  {author} {\bibinfo {author} {\bibfnamefont {M.}~\bibnamefont
			{Maggiore}},\ }\href@noop {} {\emph {\bibinfo {title} {Gravitational Waves:
				Volume 1: Theory and Experiments}}},\ Vol.~\bibinfo {volume} {1}\ (\bibinfo
	{publisher} {Oxford university press},\ \bibinfo {year} {2008})\BibitemShut
	{NoStop}%
	\bibitem [{\citenamefont {Sab{\'\i}n}\ \emph {et~al.}(2016)\citenamefont
		{Sab{\'\i}n}, \citenamefont {Kohlrus}, \citenamefont {Bruschi},\ and\
		\citenamefont {Fuentes}}]{GWDetectorThermal}%
	\BibitemOpen
	\bibfield  {author} {\bibinfo {author} {\bibfnamefont {C.}~\bibnamefont
			{Sab{\'\i}n}}, \bibinfo {author} {\bibfnamefont {J.}~\bibnamefont {Kohlrus}},
		\bibinfo {author} {\bibfnamefont {D.~E.}\ \bibnamefont {Bruschi}}, \ and\
		\bibinfo {author} {\bibfnamefont {I.}~\bibnamefont {Fuentes}},\ }\href
	{http://epjquantumtechnology.springeropen.com/articles/10.1140/epjqt/s40507-016-0046-4}
	{\bibfield  {journal} {\bibinfo  {journal} {EPJ Quantum Technology}\ }\textbf
		{\bibinfo {volume} {3}},\ \bibinfo {pages} {8} (\bibinfo {year}
		{2016})}\BibitemShut {NoStop}%
	\bibitem [{\citenamefont {Pitaevskii}\ and\ \citenamefont
		{Stringari}(2003)}]{PitaevskiiBook}%
	\BibitemOpen
	\bibfield  {author} {\bibinfo {author} {\bibfnamefont {L.}~\bibnamefont
			{Pitaevskii}}\ and\ \bibinfo {author} {\bibfnamefont {S.}~\bibnamefont
			{Stringari}},\ }\href@noop {} {\emph {\bibinfo {title} {Bose-Einstein
				Condensation}}}\ (\bibinfo  {publisher} {Oxford University Press},\ \bibinfo
	{year} {2003})\BibitemShut {NoStop}%
	\bibitem [{\citenamefont {Landau}\ and\ \citenamefont
		{Lifshitz}(1987)}]{LandauLifshitzQM}%
	\BibitemOpen
	\bibfield  {author} {\bibinfo {author} {\bibfnamefont {L.~D.}\ \bibnamefont
			{Landau}}\ and\ \bibinfo {author} {\bibfnamefont {E.~M.}\ \bibnamefont
			{Lifshitz}},\ }\href@noop {} {\emph {\bibinfo {title} {Quantum Mechanics}}},\
	\bibinfo {edition} {third edition}\ ed.\ (\bibinfo  {publisher} {Pregamon,
		Oxford},\ \bibinfo {year} {1987})\BibitemShut {NoStop}%
	\bibitem [{\citenamefont {Li}\ \emph {et~al.}(2014)\citenamefont {Li},
		\citenamefont {Yuan}, \citenamefont {Ou},\ and\ \citenamefont
		{Zhang}}]{Li_2014}%
	\BibitemOpen
	\bibfield  {author} {\bibinfo {author} {\bibfnamefont {D.}~\bibnamefont
			{Li}}, \bibinfo {author} {\bibfnamefont {C.-H.}\ \bibnamefont {Yuan}},
		\bibinfo {author} {\bibfnamefont {Z.~Y.}\ \bibnamefont {Ou}}, \ and\ \bibinfo
		{author} {\bibfnamefont {W.}~\bibnamefont {Zhang}},\ }\href {\doibase
		10.1088/1367-2630/16/7/073020} {\bibfield  {journal} {\bibinfo  {journal}
			{New Journal of Physics}\ }\textbf {\bibinfo {volume} {16}},\ \bibinfo
		{pages} {073020} (\bibinfo {year} {2014})}\BibitemShut {NoStop}%
	\bibitem [{\citenamefont {Anderson}\ \emph {et~al.}(2017)\citenamefont
		{Anderson}, \citenamefont {Gupta}, \citenamefont {Schmittberger},
		\citenamefont {Horrom}, \citenamefont {Hermann-Avigliano}, \citenamefont
		{Jones},\ and\ \citenamefont {Lett}}]{TruncatedSU11}%
	\BibitemOpen
	\bibfield  {author} {\bibinfo {author} {\bibfnamefont {B.~E.}\ \bibnamefont
			{Anderson}}, \bibinfo {author} {\bibfnamefont {P.}~\bibnamefont {Gupta}},
		\bibinfo {author} {\bibfnamefont {B.~L.}\ \bibnamefont {Schmittberger}},
		\bibinfo {author} {\bibfnamefont {T.}~\bibnamefont {Horrom}}, \bibinfo
		{author} {\bibfnamefont {C.}~\bibnamefont {Hermann-Avigliano}}, \bibinfo
		{author} {\bibfnamefont {K.~M.}\ \bibnamefont {Jones}}, \ and\ \bibinfo
		{author} {\bibfnamefont {P.~D.}\ \bibnamefont {Lett}},\ }\href {\doibase
		10.1364/OPTICA.4.000752} {\bibfield  {journal} {\bibinfo  {journal} {Optica}\
		}\textbf {\bibinfo {volume} {4}},\ \bibinfo {pages} {752} (\bibinfo {year}
		{2017})}\BibitemShut {NoStop}%
	\bibitem [{\citenamefont {Gerry}\ \emph {et~al.}(2005)\citenamefont {Gerry},
		\citenamefont {Knight},\ and\ \citenamefont {Knight}}]{KnightBook}%
	\BibitemOpen
	\bibfield  {author} {\bibinfo {author} {\bibfnamefont {C.}~\bibnamefont
			{Gerry}}, \bibinfo {author} {\bibfnamefont {P.}~\bibnamefont {Knight}}, \
		and\ \bibinfo {author} {\bibfnamefont {P.~L.}\ \bibnamefont {Knight}},\
	}\href@noop {} {\emph {\bibinfo {title} {Introductory quantum optics}}}\
	(\bibinfo  {publisher} {Cambridge University Press},\ \bibinfo {year}
	{2005})\BibitemShut {NoStop}%
	\bibitem [{\citenamefont {Birrell}\ and\ \citenamefont
		{Davies}(1982)}]{BirrelandDavies}%
	\BibitemOpen
	\bibfield  {author} {\bibinfo {author} {\bibfnamefont {N.~D.}\ \bibnamefont
			{Birrell}}\ and\ \bibinfo {author} {\bibfnamefont {P.~C.~W.}\ \bibnamefont
			{Davies}},\ }\href@noop {} {\emph {\bibinfo {title} {Quantum Fields in Curved
				Space}}}\ (\bibinfo  {publisher} {Cambridge University Press},\ \bibinfo
	{year} {1982})\BibitemShut {NoStop}%
	\bibitem [{\citenamefont {Kohlrus}(2019)}]{JanThesis}%
	\BibitemOpen
	\bibfield  {author} {\bibinfo {author} {\bibfnamefont {J.}~\bibnamefont
			{Kohlrus}},\ }\emph {\bibinfo {title} {{Quantum communication and metrology
				in curved spacetime}}},\ \href@noop {} {Ph.D. thesis},\ \bibinfo  {school}
	{University of Nottingham} (\bibinfo {year} {2019})\BibitemShut {NoStop}%
	\bibitem [{\citenamefont {Howl}\ \emph {et~al.}(2019)\citenamefont {Howl},
		\citenamefont {Fuentes} \emph {et~al.}}]{GWSeismic}%
	\BibitemOpen
	\bibfield  {author} {\bibinfo {author} {\bibfnamefont {R.}~\bibnamefont
			{Howl}}, \bibinfo {author} {\bibfnamefont {I.}~\bibnamefont {Fuentes}},
		\emph {et~al.},\ }\href@noop {} {\bibfield  {journal} {\bibinfo  {journal}
			{in preparation}\ } (\bibinfo {year} {2019})}\BibitemShut {NoStop}%
\end{thebibliography}

%

\end{document}